\documentclass[%
 reprint,preprintnumbers,
nofootinbib,
 amsmath,amssymb,
 aps
prb,
]{revtex4-1}
\pdfoutput=1
\usepackage{graphicx}
\usepackage[utf8]{inputenc}
\usepackage{flushend}
\usepackage{dcolumn}
\usepackage{bm}
\usepackage{balance}

\usepackage[normalem]{ulem}
\usepackage[colorlinks = true,
            linkcolor = blue,
            urlcolor  = blue,
            citecolor = blue,
            anchorcolor = blue]{hyperref}
\usepackage{verbatim}
\usepackage{color,ulem}
\usepackage[english]{babel}
\newcommand\new[1]{\textcolor{red}{\bf #1}}

\usepackage{MnSymbol,wasysym}

\usepackage[utf8]{inputenc}
\input Starburst.fd
\newcommand*\initfamily{\usefont{U}{Starburst}{xl}{n}}\initfamily

\newcommand{\beq}{\begin{eqnarray}}
\newcommand{\eeq}{\end{eqnarray}}
\usepackage{amsmath}
\usepackage{tikz}
\usetikzlibrary{decorations.pathmorphing}
\usetikzlibrary{shapes.misc}
\tikzset{cross/.style={cross out, draw=black, minimum size=8*(#1-\pgflinewidth), inner sep=0pt, outer sep=0pt},
cross/.default={1pt}}
\usetikzlibrary{patterns,math}

\def\be{\begin{equation}}
\def\ee{\end{equation}}
\def\bea{\begin{eqnarray}}
\def\eea{\end{eqnarray}}
\newcommand{\dd}{\mathrm{d}}
\definecolor{applegreen}{rgb}{0.55, 0.71, 0.0}

\begin{document}

\preprint{IFT-UAM/CSIC-23-85}

\title{\Large Inability of linear axion holographic Gubser-Rocha model to capture all the transport anomalies of strange metals}
\author{Yongjun Ahn$^{1,2}$}\email{yongjunahn@sjtu.edu.cn}
\author{Matteo Baggioli$^{1,2}$}\email{b.matteo@sjtu.edu.cn}
\author{Hyun-Sik Jeong$^{3,4}$}\email{hyunsik.jeong@csic.es}
\author{Keun-Young Kim$^{5,6}$}\email{fortoe@gist.ac.kr}

\affiliation{$^{1}$Wilczek Quantum Center, School of Physics and Astronomy, Shanghai Jiao Tong University, Shanghai 200240, China}
\affiliation{$^{2}$Shanghai Research Center for Quantum Sciences, Shanghai 201315, China}
\affiliation{$^3$Instituto de F\'isica Te\'orica UAM/CSIC, Calle Nicol\'as Cabrera 13-15, 28049 Madrid, Spain}
\affiliation{$^4$Departamento de F\'isica Te\'orica, Universidad Aut{\'o}noma de Madrid, Campus de Cantoblanco, 28049 Madrid, Spain}
\affiliation{$^5$Department of Physics and Photon Science, Gwangju Institute of Science and Technology,
123 Cheomdan-gwagiro, Gwangju 61005, Korea}
\affiliation{$^6$Research Center for Photon Science Technology, Gwangju Institute of Science and Technology,
123 Cheomdan-gwagiro, Gwangju 61005, Korea}

\begin{abstract}  
In the last decade, motivated by the concept of Planckian relaxation and the possible existence of a quantum critical point in cuprate materials, holographic techniques have been extensively used to tackle the problem of strange metals and high-$T_c$ superconductors. Among the various setups, the linear axion Gubser-Rocha model has often been considered as a promising holographic model for strange metals since endowed with the famous linear in $T$ resistivity property. As fiercely advocated by Phil Anderson, beyond $T$-linear resistivity, there are several additional anomalies unique to the strange metal phase, as for example a Fermi liquid like Hall angle -- the famous problem of the two relaxation scales. In this short note, we show that the linear axion holographic Gubser-Rocha model, which presents a single momentum relaxation time, fails in this respect and therefore is not able to capture the transport phenomenology of strange metals. We prove our statement by means of a direct numerical computation, a previously demonstrated scaling analysis and also a hydrodynamic argument. Finally, we conclude with an optimistic discussion on the possible improvements and generalizations which could lead to a holographic model for strange metals in all their glory.
\end{abstract}
\maketitle
\section*{Introduction}
Tackling condensed-matter problems with ``string theory'' is becoming a dedicated profession \cite{aps}. In the last decade, holography, or the gauge-gravity duality, has emerged as a promising tool to study strongly-coupled condensed matter systems \cite{zaanen_liu_sun_schalm_2015,hartnoll2018holographic,baggioli2019applied,Zaanen:2021llz}. Without any doubt, the central and original motivation for the so-called AdS-CMT program has always been the understanding of the strange metal phase and the related high-$T_c$ superconductivity in cuprate materials and other strongly-correlated systems \cite{10.1063/PT.3.1616,Faulkner:2010da}. In particular, the peculiar linear in $T$ resistivity of strange metals \cite{hussey2004universality}, and non-Fermi liquid physics in general, have always been recognized as a fundamental step in this journey and they have been chased using a large variety of models and techniques, \textit{e.g.},  \cite{PhysRevD.83.065029,Cubrovic:2009ye,Hartnoll:2009ns,Zaanen:2018edk,https://doi.org/10.1002/prop.201100030,Horowitz2013,Kiritsis:2015yna,Lucas:2015vna,Doucot:2020fvy,Cremonini:2018kla,Gan:2018utc,Blauvelt:2017koq,Amoretti:2016cad,Lee:2010ii,Pal:2010sx,Kim:2010zq,Davison:2013txa,Jeong:2018tua,Ahn:2019lrh,Balm:2022bju,Samanta:2022rkx}. 

The most famous and celebrated holographic model exhibiting linear in $T$ resistivity is the so-called Gubser-Rocha model \cite{PhysRevD.81.046001} which is a particular example of a larger class of models, known as Einstein-Maxwell-Dilaton (EMD) theories, early recognized as promising holographic effective models for condensed matter systems \cite{Charmousis:2010zz,Gouteraux:2014hca}. In this concrete setup, this strange metal resistivity is a direct consequence of the nature of the IR scale invariant fixed point, which falls in the class of the so-called semi-local quantum liquids \cite{Iqbal:2011in}. In particular, in the Gubser-Rocha model, the linear in $T$ resistivity naturally emerges from an IR fixed point in which both the hyperscaling and Lifshitz exponents, $\theta,z$, are sent to infinity by keeping their ratio negative and equal to $\theta/z=-1$. In this case, also the heat capacity scales linearly with temperature as for the Sommerfeld model \cite{Davison:2013txa}. More in general, linear in $T$ resistivity has often been associated to the properties of a specific geometry, known as AdS$_2$ \cite{PhysRevD.83.125002}, which bares important connections to other condensed matter models for strange metals such as the SYK model \cite{RevModPhys.94.035004}.

A famous critique by Phil Anderson \cite{10.1063/PT.3.1929}, and a series of recent comments \cite{Khveshchenko:2020pma,Khveshchenko:2016kmo,Khveshchenko:2015xea,Khveshchenko:2014nka,Khveshchenko:2012yt}, pointed out that holography celebrated its win too early. In particular, the concern refers to the fact that linear in $T$ resistivity is only one piece of a larger puzzle, which includes a long list of transport anomalies peculiar to strange metals \cite{Phillips:2022nxs}.

The first step towards reaching this larger picture is the observation that, despite the longitudinal resistivity $\rho_{xx}$ is ``strange'', and possibly connected to the emergence of a Planckian relaxation timescale $\rho_{xx}\sim \tau_{tr}^{-1}\sim T$ \cite{Hartnoll:2021ydi}, the Hall angle $\cot \Theta_H$ is Fermi liquid like \cite{PhysRevLett.67.2088,PhysRevLett.75.1391}, and not exhibiting the same linear in $T$ relaxation rate but rather scaling as $\cot \Theta_H \sim \tau_h^{-1}\sim T^2$. This is the famous story of the two relaxation scales suggested long time ago by Anderson \cite{PhysRevLett.67.2092}. The route to victory then necessarily includes a holographic realization of this two-scale scenario. 

A major development in the direction of reproducing, and possibly explaining, the two-scale behavior of strange metals has been proposed in \cite{Blake:2014yla}. There, using standard holographic techniques \cite{Donos:2014cya}, it has been noticed that, because of the absence of Galilean invariance, the expression for the electric DC conductivity splits into two terms:
\begin{equation}
    \sigma_{DC}=\sigma_0+\tilde \sigma,
\end{equation}
which sums up following an ``inverse-Matthiessen'' law. The first term $\sigma_0$ is forbidden in Galilean invariant systems and represents a ``charge-conjugation symmetric'' contribution (which can nevertheless depend on the total charge carrier density in some holographic models \cite{Baggioli:2016oju}). In general, $\sigma_0$, is not equal to the incoherent conductivity \cite{Blake:2015epa,Davison:2015bea}, which on the other hand corresponds to processes which do not overlap with the momentum operator \cite{hartnoll2018holographic}. The second term $\tilde \sigma$ is the contribution to the electric current from any net charge carrier density and it is strongly sensitive to the mechanisms of momentum relaxation.
Importantly, the main observation of \cite{Blake:2014yla} is that only the second term contributes to the Hall angle, which ultimately takes the form:
\begin{equation}
    \tan \Theta_H\sim \frac{B}{n} \tilde \sigma.\label{hall}
\end{equation}
This structure somehow reflects directly the two-scale scenario proposed by Anderson \cite{PhysRevLett.67.2092} and might provide an evident solution to the strange metal problem just by assuming:
\begin{equation}
    \sigma_0 \propto T^{-1},\qquad \tilde \sigma \propto T^{-2},\qquad n=\text{const}.\label{la1}
\end{equation}
This proposal resonates with the possible existence of a quantum critical point in the cuprates \cite{Marel2003}, the scaling analyses for Lifshitz-Hyperscaling critical points \cite{Hartnoll:2015sea,Karch:2015zqd} and the idea of a Planckian bound for charge diffusivity proposed by Hartnoll \cite{Hartnoll2015}. Unfortunately, this last conjecture has been proven not to be universal for the case of charge diffusivity \cite{Baggioli:2016pia}, and it cannot therefore be the origin of the scaling in Eq.\eqref{la1}. The possibility of relating a bound on Goldstone diffusivity to the linear in $T$ electric resistivity has also been discussed in the context of charge density waves where the semi-local IR fixed point still controls the scaling of $\sigma_0$ \cite{PhysRevLett.120.171603}.

As a remark, notice that the possible solution presented in Eq.\eqref{la1} is in tension with other suggestions and observations that attribute the linear in $T$ resistivity to a Drude-like term \cite{Davison:2013txa,PhysRevB.106.054515}. Those different scenarios are equivalent to assuming that $\tilde \sigma \propto T^{-1}$ and will incur in the problem that also $\cot \Theta_H \propto T$, contradicting the experimental results. Notice also that trying to explain the linear in $T$ resistivity with simple momentum relaxation physics would inevitably face the problem of explaining the universality of the coefficient $\alpha$ in $\rho_{xx}\propto \alpha T$, which is highly insensitive to the details of the material \cite{doi:10.1126/science.1227612} and even to the increase of disorder induced by irradiation \cite{PhysRevLett.91.047001}. As we will discuss, the same problem is likely to arise in the proposals which use charge density wave dynamics to this end \cite{PhysRevLett.128.141601,RevModPhys.95.011001}, since the momentum relaxation term and the CDW contributions appear always in the same combination (as noted explicitly in \cite{Amoretti:2019buu}), not giving therefore any extra freedom in the game.

Back to our focus, let us consider the holographic Gubser-Rocha model \cite{PhysRevD.81.046001}, which falls into the class I of the classification of the EMD models \cite{Gouteraux:2014hca}. In this model, the linear in $T$ resistivity comes from both terms in the DC conductivity, $\sigma_0$ and $\tilde \sigma$, scaling as $\sim T^{-1}$. Given the previous arguments, this seems already a dead end since, according to Eq.\eqref{hall}, it would imply a Hall angle scaling exactly as the longitudinal resistivity, not as in strange metals, if $n$ is temperature independent. On top of that, it has been demonstrated, and often forgot, that the transport properties of strange metals cannot be achieved in homogeneous holographic EMD models without violating the null energy condition \cite{Amoretti:2016cad}. The Gubser-Rocha model is a specific example of the models discussed in \cite{Amoretti:2016cad}. As we will explicitly show in this work, at least in its simplest formulation (\textit{e.g.}, linear Maxwell action, no explicit lattices and single momentum relaxation time), the holographic Gubser-Rocha model cannot work.

Let us mention that apart from DC thermoelectric transport, the holographic linear axion Gubser-Rocha model has been investigated in several other directions including optical transport, current-current spectral weight, fermionic spectral function and superconducting instability \cite{Anantua:2012nj,Jeong:2021wiu,Balm:2022bju,Wu:2011cy,Jeong:2019zab}. Nevertheless, to the best of our knowledge, a clear connection with strange metal physics has not been revealed for the aforementioned quantities.

The failure of the classical holographic EMD models in reproducing the transport anomalies of strange metals has been recognized in many works which tried to extend and generalize the EMD models for this purpose \cite{Kiritsis:2016cpm,Cremonini:2018kla,Gan:2018utc,Blauvelt:2017koq,Cremonini:2017qwq,Lee:2010ii,Pal:2010sx,Balm:2022bju}. Unfortunately, we are not aware of any holographic model which is able to reproduce the scaling of the electric resistivity and Hall angle in the cuprates in a large range of temperatures without resorting to the unphysical probe limit, to fine-tuning or without violating the null energy condition in the gravitational bulk. 

Despite the message that we want to deliver is rather simple, and maybe not so new, we believe that, especially for the condensed matter community facing the increasing usage of holographic models, it is important to clarify this point. Finally, and differently from other nihilist discussions in the literature \cite{Khveshchenko:2020pma,Khveshchenko:2016kmo,Khveshchenko:2015xea,Khveshchenko:2014nka,Khveshchenko:2012yt}, we would like to transmit a positive and constructive message and in particular convey the idea that the failure of the simple Gubser-Rocha model is not a failure of holography and its methods, that have been revealed to be useful and relevant in many directions, but rather an appeal for improvements and generalizations in order to achieve a description and an understanding of the physics of strange metals. Two particular promising avenues in this direction are the modification of the Maxwell term into non-linear extensions \cite{Baggioli:2016oju}, \textit{e.g.} DBI \cite{Kiritsis:2016cpm,Cremonini:2018kla,Blauvelt:2017koq}, the consideration of more general and complicated mechanisms for momentum relaxation \cite{PhysRevLett.114.251602,Baggioli:2021xuv}, and the introduction of a bona fide periodic lattice \cite{Balm:2022bju} into the EMD models. More work needs to be done.

Before moving to the bulk of the manuscript, let us make an important remark. Our definition of strange metal follows the general principles outlined by P. Anderson in \cite{10.1063/PT.3.1929}. According to this specification, a strange metal is \textit{not} equivalent to a metal solely exhibiting linear in $T$ resistivity (at low enough temperatures), as sometimes used in the literature.

The manuscript is organized as follows.
In Section \ref{sec1}, we briefly present the holographic Gubser-Rocha model; in Section \ref{sec2}, we explicitly prove that this holographic model is not able to recover the scaling of the resistivity and the Hall angle in strange metals; in Section \ref{sec3}, we show that the same conclusion could have been obtained just from a scaling analysis of the IR fixed point or from hydrodynamics; and finally, in Section \ref{sec4}, we discuss several possibilities to overcome this failure and generalize the holographic model to capture the physics of strange metals and conclude with some general remarks.

\section{The holographic Gubser-Rocha model}\label{sec1}
We study a $3+1$ dimensional holographic Einstein-Maxwell-Dilaton-Axion model:
\begin{equation}\label{GRACTION}
\begin{split}
S = & \, S_{\text{GR}} + S_{\text{axion}} =  \int \dd^4x\sqrt{-g}\left( \mathcal{L}_1 + \mathcal{L}_2   \right) \,, \\ 
& \mathcal{L}_{\text{GR}} = R-\frac{1}{4} e^\phi F^2 -\frac{3}{2}(\partial{\phi})^2+ 6\cosh \phi  \,, \\
& \mathcal{L}_{\text{axion}} = -\frac{1}{2}\sum_{I=1}^{2}(\partial \psi_{I})^2 \,,
\end{split}
\end{equation}
where we set the gravitational constant to $16 \pi G_N = 1$ and the AdS radius is chosen to be unity.
The first contribution, $S_{\text{GR}}$, corresponds to the original Gubser-Rocha model~\cite{PhysRevD.81.046001}. Here, we introduce a gauge field $A_\mu$ with its field strength $F=\dd A$, and a scalar field $\phi$ -- the dilaton.
The second term $S_{\text{axion}}$, written in terms of the axion fields $\psi_{I}$, is introduced to break isotropically translational invariance so that the resistivity becomes finite \cite{Baggioli:2021xuv}. By construction, this term will introduce a single momentum relaxation time in the dual field theory.

For the background solutions, we consider the following ansatz
\begin{equation}\label{ANSATZGR}
\begin{split}
&\dd s^2 = \frac{1}{z^2} \Bigg[-(1-z)U(z) \dd t^2+\frac{\dd z^2}{(1-z)U(z)} \\
& \qquad\qquad\,\,\, +V(z)\dd x^2 +V(z) \dd y^2 \Bigg] \,, \\
&A=(1-{z})a({z}) \dd {t} - \frac{B}{2} y \, \dd x + \frac{B}{2} x  \, \dd y \,, \\
&\phi=\frac{1}{2} \log[1+{z}\,\varphi({z})] \,, \quad \psi_{1}=k \, {x} \,, \quad \psi_{2}=k \, {y} \,,
\end{split}
\end{equation}
where $B$ is the external magnetic field and $k$ controls the strength of momentum relaxation. Here $U, V, a$ and $\varphi$ are functions of the holographic bulk coordinate $z$. The AdS boundary is located at $z=0$ and the horizon is at $z=1$: in order to ensure the asymptotic AdS boundary, all the functions in the metric are required to satisfy $U(0)=V(0)=1$.
Furthermore, one can also expand the gauge field $A_t$ near the boundary and find the chemical potential $\mu$ and the density $n$: $A_t \approx \mu - n z$. The other thermodynamic quantities such as temperature ($T$) and entropy density ($s$) are evaluated as horizon quantities using $T=U(1)/(4\pi)$ and $s=4\pi V(1)$, respectively.

For $B=0$, the model Eq. \eqref{GRACTION} allows an analytic solution which is given by
\begin{equation}\label{}
\begin{split}
U({z})&=\frac{1+(1+3 Q){z}+{z}^2(1+3 Q(1+Q)-\frac{1}{2}k^{2})}{(1+Q {z})^{3/2}} \,, \\
V({z})&=(1+Q {z})^{3/2} \,, \\
a({z})&=\frac{\sqrt{3 Q(1+Q)\left(1-\frac{k^2}{2(1+Q)^2} \right)}}{1+Q {z}} \,, \quad \varphi({z})=Q \,.
\end{split}
\end{equation}
However, in the case of a finite magnetic field, we may need to employ numerical methods and construct the solutions numerically. In this work, we fix the chemical potential and express all physical quantities in terms of the three dimensionless combinations $T/\mu, k/\mu, B/\mu^2$.

\section{Explicit proof of the absence of strange metal phenomenology}\label{sec2}
In order to proceed with the study of the transport coefficients, we employ the results of \cite{Blake:2014yla} (see also \cite{Amoretti:2015gna,Blake:2015ina}) and express the DC conductivities in terms of horizon data. This analysis yields
\begin{align}\label{DCFORHORI}
\begin{split}
\sigma_{xx} &= \frac{V k^2 \left( n^2 +B^2 e^{2\phi} + V k^2 e^{\phi} \right)}{B^2 n^2 + \left( B^2 e^{\phi} + V k^2 \right)^2} \Bigg|_{z=1} \,, \\
\sigma_{xy} &= \frac{B n \left( n^2 +B^2 e^{2\phi} + 2V k^2 e^{\phi} \right)}{B^2 n^2 + \left( B^2 e^{\phi} + V k^2 \right)^2} \Bigg|_{z=1} \,.
\end{split} 
\end{align}
From these expressions, we can define the longitudinal electric resistivity:
\begin{equation}
    \rho_{xx}=\frac{\sigma_{xx}}{\sigma_{xx}^2+\sigma_{xy}^2},
\end{equation}
and the Hall angle as:
\begin{equation}
    \cot \Theta_H=\frac{\sigma_{xx}}{\sigma_{xy}}.
\end{equation}
Thus, plugging our numerical solutions Eq.\eqref{ANSATZGR} into Eq.\eqref{DCFORHORI}, we can investigate the temperature dependence of the resistivity and the Hall angle as a function of $k/\mu$ and $B/\mu^2$.
\begin{figure}
    \centering
    \includegraphics[width=\linewidth]{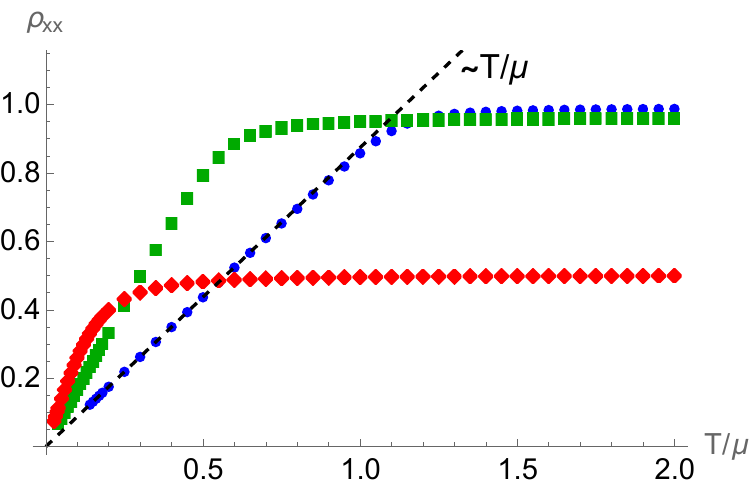}

\vspace{0.3cm}

    \includegraphics[width=\linewidth]{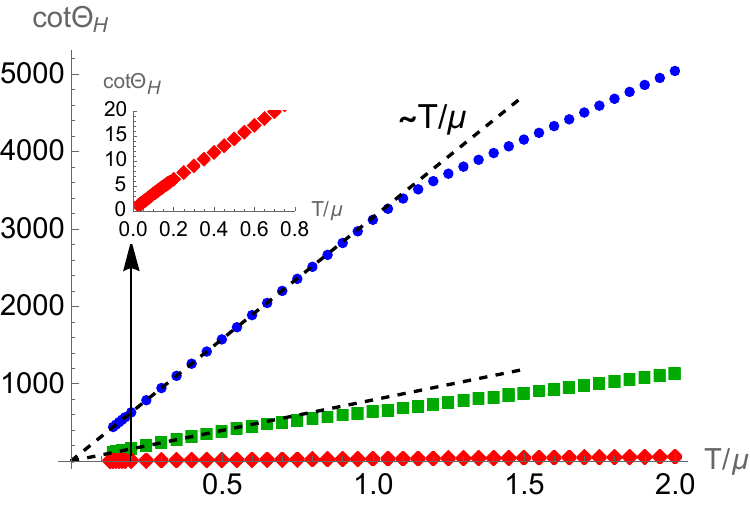}
    \caption{\textbf{Top}: The temperature dependence of the electric resistivity at various momentum relaxation rates. For each color $k/\mu$: $(\color{red}Red\color{black}, \color{applegreen}Green\color{black}, \color{blue}Blue\color{black})=(1, 5, 10)$. $B/\mu^2=1/10$. \textbf{Bottom}: The temperature dependence of the Hall angle at various momentum relaxation rates. For each color $k/\mu$: $(\color{red}Red\color{black}, \color{applegreen}Green\color{black}, \color{blue}Blue\color{black})=(1, 5, 10)$. $B/\mu^2=1/10$.}
    \label{fig:1}
\end{figure}

Fig. \ref{fig:1} shows the numerical results for the temperature dependence of the electric resistivity and the Hall angle at $B/\mu^2=1/10$, and various momentum relaxation rates. The momentum relaxation rates are denoted by colors: $k/\mu = (\color{red}\mathbf{1}\color{black}, \color{applegreen}\mathbf{5}\color{black}, \color{blue}\mathbf{10}\color{black})$. At low temperature, both the resistivity and the Hall angle exhibit linear in $T$ behavior. As the momentum relaxation increases, the low-temperature behavior extends to higher-values of the temperature. In other words, the properties of the IR fixed point extend to higher energy as already observed in \cite{Jeong:2018tua,Ahn:2019lrh}. This is reminiscent of the concept of quantum critical region, in which the properties of the quantum critical point extends to higher temperature.\\

\begin{figure}
    \centering
    \includegraphics[width=\linewidth]{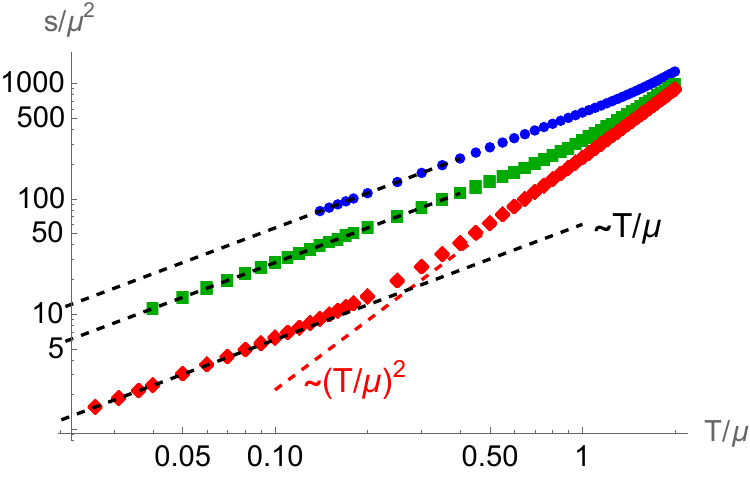}

\vspace{0.3cm}

    \includegraphics[width=\linewidth]{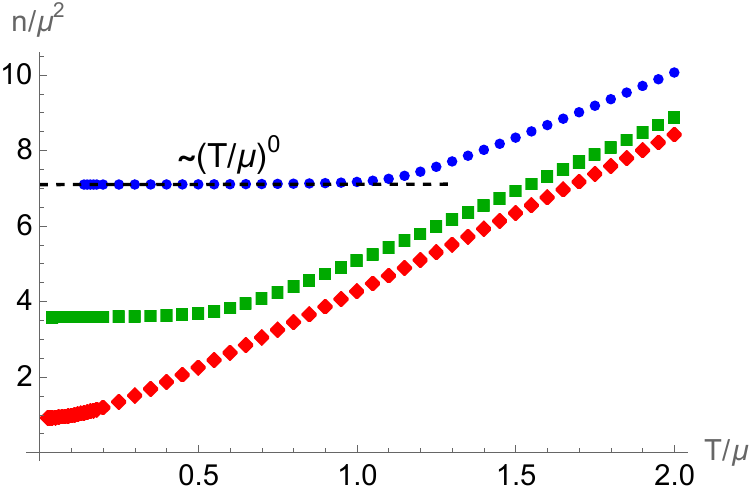}
    \caption{\textbf{Top}: The temperature dependence of the entropy density at various momentum relaxation rates. For each color $k/\mu$: $(\color{red}Red\color{black}, \color{applegreen}Green\color{black}, \color{blue}Blue\color{black})=(1, 5, 10)$. $B/\mu^2=1/10$. \textbf{Bottom}: The temperature dependence of the charge carrier density} at various momentum relaxation rates. For each color $k/\mu$: $(\color{red}Red\color{black}, \color{applegreen}Green\color{black}, \color{blue}Blue\color{black})=(1, 5, 10)$. $B/\mu^2=1/10$.
    \label{fig:2}
\end{figure}

In Fig. \ref{fig:2}, a similar behavior can be observed for the entropy density $s$ and the charge carrier density $n$. As expected, and emphasized in \cite{Davison:2013txa}, the Gubser-Rocha model exhibits a linear in temperature heat capacity which extends more and more to high energy by increasing the momentum relaxation rate $k$. We notice that a linear in $T$ capacity is not a peculiar property of strange metals but rather a common feature of all ordinary metals, in which it corresponds to the electronic contribution -- the Sommerfeld formula \cite{kittel2021introduction}. Finally, the charge carrier density $n$ is temperature independent at small temperature, as reported for example in \cite{Balm:2022bju}. As we will see in the next section, this is fully consistent with the hydrodynamic theory and it is morally the reason why the resistivity and the Hall angle scale both linearly with temperature.
\section{Scaling analysis and hydrodynamics}\label{sec3}
After proving explicitly that the Gubser-Rocha model is not able to capture the transport properties of strange metals, and in particular the temperature scaling of the electric resistivity and the Hall angle, here, we want to take a step back and re-analyze our findings from a more general perspective.

As already emphasized in the introduction, the Gubser-Rocha model is a particular case of the so-called class I in holographic EMD models (see \cite{Gouteraux:2014hca} for the details of this classification). Importantly, the analysis of all the classes of EMD models presented in \cite{Gouteraux:2014hca} has been already performed analytically in \cite{Amoretti:2016cad}. There, it has been found that for the class I, to which the Gubser-Rocha model belongs, one has always;
\begin{equation}
    \rho_{xx}\propto T^\gamma\,,\quad \cot \Theta_H \propto T^\gamma\,,
\end{equation}
where $\gamma$ depends on the details of the model and $\gamma=1$ in the Gubser-Rocha model. This is obviously consistent with the proof of concept presented in the previous section. Moreover, it shows that EMD models in class I cannot capture the strange metal transport properties.\\
Importantly, in \cite{Amoretti:2016cad} it has been shown that also all the other EMD models fail in this respect. We will discuss in the next section possible solutions to this.

Now, we want to take a different perspective and consider the same problem from a hydrodynamic point of view. For simplicity, we will focus only on the case of the electric resistivity and the Hall angle. We will start by considering the most general hydrodynamic theory which has been formulated so far in this context. The details of the hydrodynamic model can be found in \cite{Amoretti:2019buu}. Here, we will simply repeat the fundamental assumptions and report the expressions relevant for our analysis. The main assumption is to start from a 2D system at finite temperature $T$ and charge carrier density $n$ in which translational symmetry is broken pseudo-spontaneously in both directions $x,y$ and an external magnetic field perpendicular to them is introduced. This hydrodynamic framework describes the low-energy effective dynamics of charge-density wave systems \cite{RevModPhys.95.011001}. Interestingly, the form of the optical conductivity takes the same form as for the hydrodynamics of a charge fluid in a periodic external potential \cite{Chagnet:2023xsl}. For our purpose, the origin of the hydrodynamic framework will not be essential.

The only important fact is that the hydrodynamic theory predicts the following form for the electric resistivity and the Hall angle
\begin{align}
&\sigma_{DC}= \sigma_0 + \tilde \sigma,\\
    &\rho_{xx}=\frac{1}{\sigma_0+\tilde \sigma}+\mathcal{O}\left(B^2\right),\label{hydro1}\\
    & \cot \Theta_H=\frac{n}{B \tilde \sigma}\,\frac{1+\frac{\sigma_0}{\tilde \sigma}}{1+2\frac{\sigma_0}{\tilde \sigma}}+\mathcal{O}\left(B\right) \,, \label{hydro2}
\end{align}
where $\tilde \sigma$ is the combination of various contributions which depend on the particular framework considered (CDW or charged fluid in a periodic lattice for example). As a concrete example, one can consider the hydrodynamic description of a charge density wave system in presence of external magnetic field (see Appendix A in \cite{Amoretti:2019buu}). There, we can immediately identify:
\begin{equation}
    \tilde \sigma =\frac{n^2}{\chi_{\pi\pi}}\,\frac{\Omega}{\Omega \,\Gamma +\omega_0^2} \,,
\end{equation}
where $\chi_{\pi\pi}$ is the momentum susceptibility (which equals the mass density in a non-relativistic system), $\Omega$ the phase relaxation rate, $\Gamma$ the momentum relaxation rate, and $\omega_0$ the pinning frequency. On the other hand, in the simplest case of a charged electronic fluid with slow momentum relaxation, one has:
\begin{equation}
    \tilde \sigma =\frac{n^2}{\chi_{\pi\pi}\, \Gamma} \,,
\end{equation}
which is simply the well-known Drude model expression. Finally, a decomposition as in Eq.\eqref{hydro1} for the electric conductivity holds also in presence of inhomogeneous lattices \cite{Balm:2022bju,Chagnet:2023xsl}, where the expression for $\tilde \sigma$ becomes more cumbersome and depends explicitly on the lattice wave-vector.

Given Eqs.\eqref{hydro1}-\eqref{hydro2}, we would now like to ask agnostically how the scalings of the strange metals could be realized in this scenario. In order to make this analysis slightly more general, and to account also for materials in which the Hall angle is not exactly quadratic (see Table 1 in \cite{Phillips:2022nxs}), our goal will be to obtain the following situation:
\begin{equation}
    \rho_{xx}\propto T\,,\qquad \cot \Theta_H \propto T^\beta\,,
\end{equation}
with $1<\beta \leq 2$ (where $\beta=2$ corresponds to a perfectly quadratic Hall angle).

In order to perform this analysis, we first assume a generic temperature scaling for all the quantities entering in the expressions for the resistivity and the Hall angle, $\sigma_0, \tilde{\sigma}, n$, and $B$ as
\begin{equation}
    \label{scaling1}
    \sigma_0 \sim T^{-{x_1}}\,, \quad \tilde{\sigma} \sim T^{-{x_2}}\,, \quad n \sim T^{{x_3}}\,, \quad B \sim T^0\,.
\end{equation}
At this point, we will only assume that $x_1,x_2>0$ in order to have a metallic system.
Under the assumptions presented in Eq.~\eqref{scaling1}, the temperature scaling of the Hall angle $\beta$ is given by:
\begin{equation}
    \beta=x_2+x_3.
\end{equation}
We notice that the knowledge of these three scalings themselves is not enough to make predictions on the finite frequency counterpart of the above conductivities. Indeed, as seen in \cite{Amoretti:2019buu} for the case of pinned charge density waves, fixing the above scalings is in general not even enough to derive the temperature behavior of the leading order hydrodynamic coefficients since $\tilde \sigma$ involves a complex combination of many of them.\\

Let us now list the various options available.\\

\noindent \textbf{Option 1: constant density and linear in $T$ from charge-conjugation symmetric conductivity.} The first option is to assume that the charge carrier density is temperature independent and the linear in $T$ resistivity comes from the dissipative part of the conductivity $\tilde \sigma$. This option includes the proposal of \cite{Davison:2013txa} that linear in $T$ comes from the momentum relaxation rate, but also the idea of \cite{PhysRevLett.128.141601} in which linear in $T$ comes from the Goldstone diffusivity and charge density wave physics. This option corresponds to having $x_3=0$ together with $x_2=1$ and $x_1\leq1$. Given these values, the scaling of the Hall angle is equal to that in the electric resistivity since $\beta=x_3+x_2=1$, as for the holographic Gubser Rocha model just analyzed. This is in clear tension with the experimental observations. This option, and consequently the ideas in \cite{Davison:2013txa,PhysRevLett.128.141601}, are not a viable possibility to reproduce the phenomenology of strange metals.
\\

\noindent \textbf{Option 2: constant density and linear in $T$ from dissipative} physics. A second option consists in keeping the charge carrier density independent of temperature but deriving the linear in $T$ resistivity from the incoherent part of the conductivity $\sigma_0$. This corresponds to the original idea that linear in $T$ resistivity is the result of a IR quantum critical point with possibly Lifshitz and Hyperscaling features. In this case, we have that $x_3=0$ together with $x_1=1$ and $x_2\leq1$. In consequence, the scaling of the Hall angle is  always less than $1$ since $\beta=x_3+x_2\leq1$. This is again in tension with the experimental observations. Notice that this problem was solved in \cite{Blake:2014yla} by assuming that the linear in $T$ resistivity does not extend down to zero temperature but appears only above a certain energy scale $W$. In that case, by assuming $x_1=1$ and $x_2=2$ one obtains a quadratic Hall angle together with a resistivity of the form:
\begin{equation}
    \rho_{xx}\propto \frac{T^2}{W+T}\,,
\end{equation}
which is linear for $T \gg W$. Two comments are in order. First, it is not clear to us how much solid evidence there is for the linear in $T$ scaling to be valid up to zero temperature. If that is the case, then $W \rightarrow 0$, invalidating this option. On the other hand, the meaning and value of $W$ is also unclear and probably material dependent.
\\

\noindent \textbf{Option 3: temperature-dependent charge carrier density.} Given the negative results for options 1 and 2, it seems that without adding any further ingredients the only possibility is to assume the charge carrier density to be temperature dependent. Interestingly, this scenario is confirmed by the experimental fits in Bi-2201 for example \cite{Amoretti:2019buu}. Unfortunately, as explicitly emphasized in \cite{Balm:2022bju}, both Reissner-Nordstrom model and Gubser-Rocha model have $x_3=0$ and a constant in temperature charge carrier density at low temperature. By assuming $x_3\neq 0$, we can have a solution. In particular, we can extend option 1 and have:
\begin{equation}
  {\Large \smiley{}}: \qquad  x_2=1,\qquad x_3=1,\qquad x_1\leq 1;
\end{equation}
or extend option 2 and have:
\begin{equation}
   {\Large \smiley{}}: \qquad x_1=1,\qquad x_2\leq 1,\quad 1<x_3+x_2\leq 2\,.
\end{equation}
In summary, the hydrodynamic analysis suggests that a possible way out to recover the strange metal scalings (without introducing additional structures) is simply to make the charge carrier density dependent on temperature, as reported for Bi-2201 in \cite{Amoretti:2019buu}. Moreover, one could still save option 2, but only assuming that the linear in $T$ scaling of the electric resistivity appears only above a certain energy scale and does not extend down to zero temperature.

We notice that recent experimental results \cite{PhysRevB.106.054515} seem to indicate that the quantum critical conductivity $\sigma_0$ has no role in DC transport, implying a contradiction with option 1. Also, other experimental results \cite{Amoretti:2019buu} display a temperature dependent charge carrier density favoring option 3, and at odd with options 1-2. It would be useful to compare in more detail these options, and their possibly testable predictions, with the existing simulations and experiments. In this respect, it would be interesting to find a generalization of the linear axion Gubser-Rocha model to accommodate for option 3.
\section{Conclusions \& way out}\label{sec4}
In this work, we have shown that the electric resistivity and the Hall angle in the holographic linear axion Gubser-Rocha model exhibit the same linear in temperature scaling at low temperature. This implies that the linear axion Gubser-Rocha model is not a good holographic setup for strange metals, which on the contrary display different temperature scalings for those transport properties. The failure of the linear axion Gubser-Rocha model should not come as a surprise since a previous extended analysis on the EMD holographic models \cite{Amoretti:2016cad} already proved the impossibility of reproducing the strange metal scalings in a much larger class of setups. Furthermore, using a simple but general enough hydrodynamic description for strange metals, the same conclusion can be reached.

At this point, the relevant question is how to generalize and extend the holographic models to incorporate the strange metal phenomenology. A few options are available. 

The first option is to modify the Maxwell sector, responsible for the dynamics of the charge current. This can be done in two ways. One can substitute the linear Maxwell term $F_{\mu\nu}F^{\mu\nu}$ with a more general non-linear extension as for example done in \cite{Baggioli:2016oju}. A natural candidate in this direction is the DBI action which has been considered in several instances \cite{Cremonini:2018kla,Gan:2018utc,Blauvelt:2017koq,Cremonini:2017qwq,Lee:2010ii,Pal:2010sx}. The advantage of this route is that it might possibly reproduce also the $\sqrt{a_1T^2+a_2B^2}$ magneto-resistance structure \cite{Kiritsis:2016cpm} observed in certain compounds \cite{Hayes2016}. A second way is to couple directly the momentum relaxation sector with the Maxwell sector, as done in \cite{Baggioli:2016oqk,Gouteraux:2016wxj,Baggioli:2016pia}.

The second option is to modify the momentum relaxation sector beyond the original linear-axion model. This can be done for example by assuming a more general potential for the axion fields \cite{PhysRevLett.114.251602,Taylor:2014tka}. From the gravity point of view, this would correspond to consider the most general Lorentz violating massive gravity theory \cite{Alberte:2015isw} which allows for a larger freedom in the temperature dependence of the momentum relaxation rate (\textit{cfr.}, the linear axion case \cite{Davison:2013jba}). Additionally, it is important to notice\footnote{\new{We thank Koenraad Schalm for discussion and useful suggestions about this point.}} that the choice of linear axions implies a single momentum relaxation time in the dual picture. In the pursue of realizing the scenario proposed by Anderson \cite{PhysRevLett.67.2092}, it would be important to generalize the momentum relaxation sector to introduce two relaxation times. This can be probably done using an anisotropic linear axion model with different dilatonic couplings in the $x$ and $y$ directions. We leave this task for future work. Also, it would be interesting to understand if a temperature dependent charge carrier density, $n(T)$, as reported from experimental data for example in Bi-2201 \cite{Amoretti:2019buu}, could be the solution to this problem. Unfortunately, as emphasized in the previous section, the Gubser-Rocha model does not allow for this option.

More complicated escapes involve the introduction of an explicit lattice \cite{Balm:2022bju} or of explicit disorder which can modify the nature of the IR fixed point (see for example \cite{Huang:2023ihu}). One could also think about making the magnetic field relevant in the IR (see discussions in \cite{Amoretti:2016cad}). In that case, the IR structure of the theory will be strongly modified and the scaling analysis not applicable anymore, making a numerical analysis necessary. This scenario might be also useful to study in more detail the incoherent Hall conductivity recently discussed in \cite{Amoretti:2020mkp}. Finally, a last possibility is to relax more symmetries by for example introducing explicit anisotropy or considering the recent proposal for ersatz Fermi liquids of \cite{else2023holographic}.

A totally different view on the problem would be to assume that these scaling properties are not extended down to zero temperature but they do appear only above a certain, and possibly small, energy scale $W$. This was the solution proposed in \cite{Blake:2014yla}. In all honesty, we do not know how much experimental evidence in favor of this possibility exist in the literature. On the other side, we can confidently say that this scenario cannot be realized in the Gubser-Rocha model. It could be realized in other classes of EMD models, as described in \cite{Amoretti:2016cad}. It would be fruitful to investigate this further.

In summary, a holographic model for strange metals should be as simple as possible but not simpler than that. Unfortunately, the Gubser-Rocha model is simpler than that and still not the final answer to the strange metal puzzle. It would be interesting to understand how many of the alternative non-holographic options could provide a positive answer in this regard and see a similar critical analysis, as done in this work for the Gubser-Rocha model, applied to those scenarios. Curiously, Table 2 in \cite{Phillips:2022nxs} does not report this information.  

\section*{Acknowledgments}
We would like to thank Andrea Amoretti, Philip W. Phillips, and Li Li for several discussions on the topic of this work and useful comments on a preliminary version of this manuscript. We thank Koenraad Schalm, Xianhui Ge, Malte Grosche, Kamran Behnia and Hui Xing for useful comments and suggestions.
YA and MB acknowledge the support of the Shanghai Municipal Science and Technology Major Project (Grant No.2019SHZDZX01). MB acknowledges the sponsorship from the Yangyang Development Fund.
H.-S Jeong acknowledges the support of the Spanish MINECO ``Centro de Excelencia Severo Ochoa'' Programme under grant SEV-2012-0249. This work is supported through the grants CEX2020-001007-S and PID2021-123017NB-I00, funded by MCIN/AEI/10.13039/501100011033 and by ERDF A way of making Europe.
KK was supported by the Basic Science Research Program through the National Research Foundation of Korea (NRF) funded by the Ministry of Science, ICT $\&$ Future Planning (NRF-2021R1A2C1006791) and GIST Research Institute(GRI) grant funded by the GIST in 2023. KK was also supported by Creation
of the Quantum Information Science R\&D Ecosystem (Grant No. 2022M3H3A106307411)
through the National Research Foundation of Korea (NRF) funded by the Korean government (Ministry of Science and ICT).

\bibliographystyle{apsrev4-1}
\bibliography{holo}

\begin{thebibliography}{84}%
\makeatletter
\providecommand \@ifxundefined [1]{%
 \@ifx{#1\undefined}
}%
\providecommand \@ifnum [1]{%
 \ifnum #1\expandafter \@firstoftwo
 \else \expandafter \@secondoftwo
 \fi
}%
\providecommand \@ifx [1]{%
 \ifx #1\expandafter \@firstoftwo
 \else \expandafter \@secondoftwo
 \fi
}%
\providecommand \natexlab [1]{#1}%
\providecommand \enquote  [1]{``#1''}%
\providecommand \bibnamefont  [1]{#1}%
\providecommand \bibfnamefont [1]{#1}%
\providecommand \citenamefont [1]{#1}%
\providecommand \href@noop [0]{\@secondoftwo}%
\providecommand \href [0]{\begingroup \@sanitize@url \@href}%
\providecommand \@href[1]{\@@startlink{#1}\@@href}%
\providecommand \@@href[1]{\endgroup#1\@@endlink}%
\providecommand \@sanitize@url [0]{\catcode `\\12\catcode `\$12\catcode
  `\&12\catcode `\#12\catcode `\^12\catcode `\_12\catcode `\%12\relax}%
\providecommand \@@startlink[1]{}%
\providecommand \@@endlink[0]{}%
\providecommand \url  [0]{\begingroup\@sanitize@url \@url }%
\providecommand \@url [1]{\endgroup\@href {#1}{\urlprefix }}%
\providecommand \urlprefix  [0]{URL }%
\providecommand \Eprint [0]{\href }%
\providecommand \doibase [0]{http://dx.doi.org/}%
\providecommand \selectlanguage [0]{\@gobble}%
\providecommand \bibinfo  [0]{\@secondoftwo}%
\providecommand \bibfield  [0]{\@secondoftwo}%
\providecommand \translation [1]{[#1]}%
\providecommand \BibitemOpen [0]{}%
\providecommand \bibitemStop [0]{}%
\providecommand \bibitemNoStop [0]{.\EOS\space}%
\providecommand \EOS [0]{\spacefactor3000\relax}%
\providecommand \BibitemShut  [1]{\csname bibitem#1\endcsname}%
\let\auto@bib@innerbib\@empty
\bibitem [{\citenamefont {Schirber}(2020)}]{aps}%
  \BibitemOpen
  \bibfield  {author} {\bibinfo {author} {\bibfnamefont {M.}~\bibnamefont
  {Schirber}},\ }\href {https://physics.aps.org/articles/v13/57} {\enquote
  {\bibinfo {title} {Holographist by trade},}\ } (\bibinfo {year}
  {2020})\BibitemShut {NoStop}%
\bibitem [{\citenamefont {Zaanen}\ \emph {et~al.}(2015)\citenamefont {Zaanen},
  \citenamefont {Liu}, \citenamefont {Sun},\ and\ \citenamefont
  {Schalm}}]{zaanen_liu_sun_schalm_2015}%
  \BibitemOpen
  \bibfield  {author} {\bibinfo {author} {\bibfnamefont {J.}~\bibnamefont
  {Zaanen}}, \bibinfo {author} {\bibfnamefont {Y.}~\bibnamefont {Liu}},
  \bibinfo {author} {\bibfnamefont {Y.-W.}\ \bibnamefont {Sun}}, \ and\
  \bibinfo {author} {\bibfnamefont {K.}~\bibnamefont {Schalm}},\ }\href
  {\doibase 10.1017/CBO9781139942492} {\emph {\bibinfo {title} {Holographic
  Duality in Condensed Matter Physics}}}\ (\bibinfo  {publisher} {Cambridge
  University Press},\ \bibinfo {year} {2015})\BibitemShut {NoStop}%
\bibitem [{\citenamefont {Hartnoll}\ \emph {et~al.}(2018)\citenamefont
  {Hartnoll}, \citenamefont {Lucas},\ and\ \citenamefont
  {Sachdev}}]{hartnoll2018holographic}%
  \BibitemOpen
  \bibfield  {author} {\bibinfo {author} {\bibfnamefont {S.~A.}\ \bibnamefont
  {Hartnoll}}, \bibinfo {author} {\bibfnamefont {A.}~\bibnamefont {Lucas}}, \
  and\ \bibinfo {author} {\bibfnamefont {S.}~\bibnamefont {Sachdev}},\
  }\href@noop {} {\emph {\bibinfo {title} {Holographic quantum matter}}}\
  (\bibinfo  {publisher} {MIT press},\ \bibinfo {year} {2018})\BibitemShut
  {NoStop}%
\bibitem [{\citenamefont {Baggioli}(2019)}]{baggioli2019applied}%
  \BibitemOpen
  \bibfield  {author} {\bibinfo {author} {\bibfnamefont {M.}~\bibnamefont
  {Baggioli}},\ }\href@noop {} {\emph {\bibinfo {title} {Applied holography: a
  practical mini-course}}}\ (\bibinfo  {publisher} {Springer},\ \bibinfo {year}
  {2019})\BibitemShut {NoStop}%
\bibitem [{\citenamefont {Zaanen}(2021)}]{Zaanen:2021llz}%
  \BibitemOpen
  \bibfield  {author} {\bibinfo {author} {\bibfnamefont {J.}~\bibnamefont
  {Zaanen}},\ }\href@noop {} {\  (\bibinfo {year} {2021})},\ \Eprint
  {http://arxiv.org/abs/2110.00961} {arXiv:2110.00961 [cond-mat.str-el]}
  \BibitemShut {NoStop}%
\bibitem [{\citenamefont {Liu}(2012)}]{10.1063/PT.3.1616}%
  \BibitemOpen
  \bibfield  {author} {\bibinfo {author} {\bibfnamefont {H.}~\bibnamefont
  {Liu}},\ }\href {\doibase 10.1063/PT.3.1616} {\bibfield  {journal} {\bibinfo
  {journal} {Physics Today}\ }\textbf {\bibinfo {volume} {65}},\ \bibinfo
  {pages} {68} (\bibinfo {year} {2012})}\BibitemShut {NoStop}%
\bibitem [{\citenamefont {Faulkner}\ \emph {et~al.}(2010)\citenamefont
  {Faulkner}, \citenamefont {Iqbal}, \citenamefont {Liu}, \citenamefont
  {McGreevy},\ and\ \citenamefont {Vegh}}]{Faulkner:2010da}%
  \BibitemOpen
  \bibfield  {author} {\bibinfo {author} {\bibfnamefont {T.}~\bibnamefont
  {Faulkner}}, \bibinfo {author} {\bibfnamefont {N.}~\bibnamefont {Iqbal}},
  \bibinfo {author} {\bibfnamefont {H.}~\bibnamefont {Liu}}, \bibinfo {author}
  {\bibfnamefont {J.}~\bibnamefont {McGreevy}}, \ and\ \bibinfo {author}
  {\bibfnamefont {D.}~\bibnamefont {Vegh}},\ }\href@noop {} {\  (\bibinfo
  {year} {2010})},\ \Eprint {http://arxiv.org/abs/1003.1728} {arXiv:1003.1728
  [hep-th]} \BibitemShut {NoStop}%
\bibitem [{\citenamefont {Hussey‖}\ \emph {et~al.}(2004)\citenamefont
  {Hussey‖}, \citenamefont {Takenaka},\ and\ \citenamefont
  {Takagi}}]{hussey2004universality}%
  \BibitemOpen
  \bibfield  {author} {\bibinfo {author} {\bibfnamefont {N.}~\bibnamefont
  {Hussey‖}}, \bibinfo {author} {\bibfnamefont {K.}~\bibnamefont {Takenaka}},
  \ and\ \bibinfo {author} {\bibfnamefont {H.}~\bibnamefont {Takagi}},\
  }\href@noop {} {\bibfield  {journal} {\bibinfo  {journal} {Philosophical
  Magazine}\ }\textbf {\bibinfo {volume} {84}},\ \bibinfo {pages} {2847}
  (\bibinfo {year} {2004})}\BibitemShut {NoStop}%
\bibitem [{\citenamefont {Liu}\ \emph {et~al.}(2011)\citenamefont {Liu},
  \citenamefont {McGreevy},\ and\ \citenamefont {Vegh}}]{PhysRevD.83.065029}%
  \BibitemOpen
  \bibfield  {author} {\bibinfo {author} {\bibfnamefont {H.}~\bibnamefont
  {Liu}}, \bibinfo {author} {\bibfnamefont {J.}~\bibnamefont {McGreevy}}, \
  and\ \bibinfo {author} {\bibfnamefont {D.}~\bibnamefont {Vegh}},\ }\href
  {\doibase 10.1103/PhysRevD.83.065029} {\bibfield  {journal} {\bibinfo
  {journal} {Phys. Rev. D}\ }\textbf {\bibinfo {volume} {83}},\ \bibinfo
  {pages} {065029} (\bibinfo {year} {2011})}\BibitemShut {NoStop}%
\bibitem [{\citenamefont {Cubrovic}\ \emph {et~al.}(2009)\citenamefont
  {Cubrovic}, \citenamefont {Zaanen},\ and\ \citenamefont
  {Schalm}}]{Cubrovic:2009ye}%
  \BibitemOpen
  \bibfield  {author} {\bibinfo {author} {\bibfnamefont {M.}~\bibnamefont
  {Cubrovic}}, \bibinfo {author} {\bibfnamefont {J.}~\bibnamefont {Zaanen}}, \
  and\ \bibinfo {author} {\bibfnamefont {K.}~\bibnamefont {Schalm}},\ }\href
  {\doibase 10.1126/science.1174962} {\bibfield  {journal} {\bibinfo  {journal}
  {Science}\ }\textbf {\bibinfo {volume} {325}},\ \bibinfo {pages} {439}
  (\bibinfo {year} {2009})},\ \Eprint {http://arxiv.org/abs/0904.1993}
  {arXiv:0904.1993 [hep-th]} \BibitemShut {NoStop}%
\bibitem [{\citenamefont {Hartnoll}\ \emph {et~al.}(2010)\citenamefont
  {Hartnoll}, \citenamefont {Polchinski}, \citenamefont {Silverstein},\ and\
  \citenamefont {Tong}}]{Hartnoll:2009ns}%
  \BibitemOpen
  \bibfield  {author} {\bibinfo {author} {\bibfnamefont {S.~A.}\ \bibnamefont
  {Hartnoll}}, \bibinfo {author} {\bibfnamefont {J.}~\bibnamefont
  {Polchinski}}, \bibinfo {author} {\bibfnamefont {E.}~\bibnamefont
  {Silverstein}}, \ and\ \bibinfo {author} {\bibfnamefont {D.}~\bibnamefont
  {Tong}},\ }\href {\doibase 10.1007/JHEP04(2010)120} {\bibfield  {journal}
  {\bibinfo  {journal} {JHEP}\ }\textbf {\bibinfo {volume} {04}},\ \bibinfo
  {pages} {120} (\bibinfo {year} {2010})},\ \Eprint
  {http://arxiv.org/abs/0912.1061} {arXiv:0912.1061 [hep-th]} \BibitemShut
  {NoStop}%
\bibitem [{\citenamefont {Zaanen}(2019)}]{Zaanen:2018edk}%
  \BibitemOpen
  \bibfield  {author} {\bibinfo {author} {\bibfnamefont {J.}~\bibnamefont
  {Zaanen}},\ }\href {\doibase 10.21468/SciPostPhys.6.5.061} {\bibfield
  {journal} {\bibinfo  {journal} {SciPost Phys.}\ }\textbf {\bibinfo {volume}
  {6}},\ \bibinfo {pages} {061} (\bibinfo {year} {2019})},\ \Eprint
  {http://arxiv.org/abs/1807.10951} {arXiv:1807.10951 [cond-mat.str-el]}
  \BibitemShut {NoStop}%
\bibitem [{\citenamefont {Meyer}\ \emph {et~al.}(2011)\citenamefont {Meyer},
  \citenamefont {Goutéraux},\ and\ \citenamefont
  {Kim}}]{https://doi.org/10.1002/prop.201100030}%
  \BibitemOpen
  \bibfield  {author} {\bibinfo {author} {\bibfnamefont {R.}~\bibnamefont
  {Meyer}}, \bibinfo {author} {\bibfnamefont {B.}~\bibnamefont {Goutéraux}}, \
  and\ \bibinfo {author} {\bibfnamefont {B.}~\bibnamefont {Kim}},\ }\href
  {\doibase https://doi.org/10.1002/prop.201100030} {\bibfield  {journal}
  {\bibinfo  {journal} {Fortschritte der Physik}\ }\textbf {\bibinfo {volume}
  {59}},\ \bibinfo {pages} {741} (\bibinfo {year} {2011})}\BibitemShut
  {NoStop}%
\bibitem [{\citenamefont {Horowitz}\ and\ \citenamefont
  {Santos}(2013)}]{Horowitz2013}%
  \BibitemOpen
  \bibfield  {author} {\bibinfo {author} {\bibfnamefont {G.~T.}\ \bibnamefont
  {Horowitz}}\ and\ \bibinfo {author} {\bibfnamefont {J.~E.}\ \bibnamefont
  {Santos}},\ }\href {\doibase 10.1007/JHEP06(2013)087} {\bibfield  {journal}
  {\bibinfo  {journal} {Journal of High Energy Physics}\ }\textbf {\bibinfo
  {volume} {2013}},\ \bibinfo {pages} {87} (\bibinfo {year}
  {2013})}\BibitemShut {NoStop}%
\bibitem [{\citenamefont {Kiritsis}\ and\ \citenamefont {Pe\~na
  Benitez}(2015)}]{Kiritsis:2015yna}%
  \BibitemOpen
  \bibfield  {author} {\bibinfo {author} {\bibfnamefont {E.}~\bibnamefont
  {Kiritsis}}\ and\ \bibinfo {author} {\bibfnamefont {F.}~\bibnamefont {Pe\~na
  Benitez}},\ }\href {\doibase 10.1007/JHEP11(2015)177} {\bibfield  {journal}
  {\bibinfo  {journal} {JHEP}\ }\textbf {\bibinfo {volume} {11}},\ \bibinfo
  {pages} {177} (\bibinfo {year} {2015})},\ \Eprint
  {http://arxiv.org/abs/1507.05633} {arXiv:1507.05633 [cond-mat.str-el]}
  \BibitemShut {NoStop}%
\bibitem [{\citenamefont {Lucas}(2015)}]{Lucas:2015vna}%
  \BibitemOpen
  \bibfield  {author} {\bibinfo {author} {\bibfnamefont {A.}~\bibnamefont
  {Lucas}},\ }\href {\doibase 10.1007/JHEP03(2015)071} {\bibfield  {journal}
  {\bibinfo  {journal} {JHEP}\ }\textbf {\bibinfo {volume} {03}},\ \bibinfo
  {pages} {071} (\bibinfo {year} {2015})},\ \Eprint
  {http://arxiv.org/abs/1501.05656} {arXiv:1501.05656 [hep-th]} \BibitemShut
  {NoStop}%
\bibitem [{\citenamefont {Doucot}\ \emph {et~al.}(2021)\citenamefont {Doucot},
  \citenamefont {Mukhopadhyay}, \citenamefont {Policastro},\ and\ \citenamefont
  {Samanta}}]{Doucot:2020fvy}%
  \BibitemOpen
  \bibfield  {author} {\bibinfo {author} {\bibfnamefont {B.}~\bibnamefont
  {Doucot}}, \bibinfo {author} {\bibfnamefont {A.}~\bibnamefont
  {Mukhopadhyay}}, \bibinfo {author} {\bibfnamefont {G.}~\bibnamefont
  {Policastro}}, \ and\ \bibinfo {author} {\bibfnamefont {S.}~\bibnamefont
  {Samanta}},\ }\href {\doibase 10.1103/PhysRevD.104.L081901} {\bibfield
  {journal} {\bibinfo  {journal} {Phys. Rev. D}\ }\textbf {\bibinfo {volume}
  {104}},\ \bibinfo {pages} {L081901} (\bibinfo {year} {2021})},\ \Eprint
  {http://arxiv.org/abs/2012.15679} {arXiv:2012.15679 [hep-th]} \BibitemShut
  {NoStop}%
\bibitem [{\citenamefont {Cremonini}\ \emph {et~al.}(2019)\citenamefont
  {Cremonini}, \citenamefont {Hoover}, \citenamefont {Li},\ and\ \citenamefont
  {Waskie}}]{Cremonini:2018kla}%
  \BibitemOpen
  \bibfield  {author} {\bibinfo {author} {\bibfnamefont {S.}~\bibnamefont
  {Cremonini}}, \bibinfo {author} {\bibfnamefont {A.}~\bibnamefont {Hoover}},
  \bibinfo {author} {\bibfnamefont {L.}~\bibnamefont {Li}}, \ and\ \bibinfo
  {author} {\bibfnamefont {S.}~\bibnamefont {Waskie}},\ }\href {\doibase
  10.1103/PhysRevD.99.061901} {\bibfield  {journal} {\bibinfo  {journal} {Phys.
  Rev. D}\ }\textbf {\bibinfo {volume} {99}},\ \bibinfo {pages} {061901}
  (\bibinfo {year} {2019})},\ \Eprint {http://arxiv.org/abs/1812.01040}
  {arXiv:1812.01040 [hep-th]} \BibitemShut {NoStop}%
\bibitem [{\citenamefont {Gan}\ \emph {et~al.}(2019)\citenamefont {Gan},
  \citenamefont {Wang},\ and\ \citenamefont {Yang}}]{Gan:2018utc}%
  \BibitemOpen
  \bibfield  {author} {\bibinfo {author} {\bibfnamefont {Q.}~\bibnamefont
  {Gan}}, \bibinfo {author} {\bibfnamefont {P.}~\bibnamefont {Wang}}, \ and\
  \bibinfo {author} {\bibfnamefont {H.}~\bibnamefont {Yang}},\ }\href {\doibase
  10.1088/0253-6102/71/5/577} {\bibfield  {journal} {\bibinfo  {journal}
  {Commun. Theor. Phys.}\ }\textbf {\bibinfo {volume} {71}},\ \bibinfo {pages}
  {577} (\bibinfo {year} {2019})},\ \Eprint {http://arxiv.org/abs/1808.06158}
  {arXiv:1808.06158 [hep-th]} \BibitemShut {NoStop}%
\bibitem [{\citenamefont {Blauvelt}\ \emph {et~al.}(2018)\citenamefont
  {Blauvelt}, \citenamefont {Cremonini}, \citenamefont {Hoover}, \citenamefont
  {Li},\ and\ \citenamefont {Waskie}}]{Blauvelt:2017koq}%
  \BibitemOpen
  \bibfield  {author} {\bibinfo {author} {\bibfnamefont {E.}~\bibnamefont
  {Blauvelt}}, \bibinfo {author} {\bibfnamefont {S.}~\bibnamefont {Cremonini}},
  \bibinfo {author} {\bibfnamefont {A.}~\bibnamefont {Hoover}}, \bibinfo
  {author} {\bibfnamefont {L.}~\bibnamefont {Li}}, \ and\ \bibinfo {author}
  {\bibfnamefont {S.}~\bibnamefont {Waskie}},\ }\href {\doibase
  10.1103/PhysRevD.97.061901} {\bibfield  {journal} {\bibinfo  {journal} {Phys.
  Rev. D}\ }\textbf {\bibinfo {volume} {97}},\ \bibinfo {pages} {061901}
  (\bibinfo {year} {2018})},\ \Eprint {http://arxiv.org/abs/1710.01326}
  {arXiv:1710.01326 [hep-th]} \BibitemShut {NoStop}%
\bibitem [{\citenamefont {Amoretti}\ \emph {et~al.}(2016)\citenamefont
  {Amoretti}, \citenamefont {Baggioli}, \citenamefont {Magnoli},\ and\
  \citenamefont {Musso}}]{Amoretti:2016cad}%
  \BibitemOpen
  \bibfield  {author} {\bibinfo {author} {\bibfnamefont {A.}~\bibnamefont
  {Amoretti}}, \bibinfo {author} {\bibfnamefont {M.}~\bibnamefont {Baggioli}},
  \bibinfo {author} {\bibfnamefont {N.}~\bibnamefont {Magnoli}}, \ and\
  \bibinfo {author} {\bibfnamefont {D.}~\bibnamefont {Musso}},\ }\href
  {\doibase 10.1007/JHEP06(2016)113} {\bibfield  {journal} {\bibinfo  {journal}
  {JHEP}\ }\textbf {\bibinfo {volume} {06}},\ \bibinfo {pages} {113} (\bibinfo
  {year} {2016})},\ \Eprint {http://arxiv.org/abs/1603.03029} {arXiv:1603.03029
  [hep-th]} \BibitemShut {NoStop}%
\bibitem [{\citenamefont {Lee}\ \emph {et~al.}(2010)\citenamefont {Lee},
  \citenamefont {Pang},\ and\ \citenamefont {Park}}]{Lee:2010ii}%
  \BibitemOpen
  \bibfield  {author} {\bibinfo {author} {\bibfnamefont {B.-H.}\ \bibnamefont
  {Lee}}, \bibinfo {author} {\bibfnamefont {D.-W.}\ \bibnamefont {Pang}}, \
  and\ \bibinfo {author} {\bibfnamefont {C.}~\bibnamefont {Park}},\ }\href
  {\doibase 10.1007/JHEP07(2010)057} {\bibfield  {journal} {\bibinfo  {journal}
  {JHEP}\ }\textbf {\bibinfo {volume} {07}},\ \bibinfo {pages} {057} (\bibinfo
  {year} {2010})},\ \Eprint {http://arxiv.org/abs/1006.1719} {arXiv:1006.1719
  [hep-th]} \BibitemShut {NoStop}%
\bibitem [{\citenamefont {Pal}(2011)}]{Pal:2010sx}%
  \BibitemOpen
  \bibfield  {author} {\bibinfo {author} {\bibfnamefont {S.~S.}\ \bibnamefont
  {Pal}},\ }\href {\doibase 10.1103/PhysRevD.84.126009} {\bibfield  {journal}
  {\bibinfo  {journal} {Phys. Rev. D}\ }\textbf {\bibinfo {volume} {84}},\
  \bibinfo {pages} {126009} (\bibinfo {year} {2011})},\ \Eprint
  {http://arxiv.org/abs/1011.3117} {arXiv:1011.3117 [hep-th]} \BibitemShut
  {NoStop}%
\bibitem [{\citenamefont {Kim}\ \emph {et~al.}(2012)\citenamefont {Kim},
  \citenamefont {Kiritsis},\ and\ \citenamefont {Panagopoulos}}]{Kim:2010zq}%
  \BibitemOpen
  \bibfield  {author} {\bibinfo {author} {\bibfnamefont {B.~S.}\ \bibnamefont
  {Kim}}, \bibinfo {author} {\bibfnamefont {E.}~\bibnamefont {Kiritsis}}, \
  and\ \bibinfo {author} {\bibfnamefont {C.}~\bibnamefont {Panagopoulos}},\
  }\href {\doibase 10.1088/1367-2630/14/4/043045} {\bibfield  {journal}
  {\bibinfo  {journal} {New J. Phys.}\ }\textbf {\bibinfo {volume} {14}},\
  \bibinfo {pages} {043045} (\bibinfo {year} {2012})},\ \Eprint
  {http://arxiv.org/abs/1012.3464} {arXiv:1012.3464 [cond-mat.str-el]}
  \BibitemShut {NoStop}%
\bibitem [{\citenamefont {Davison}\ \emph {et~al.}(2014)\citenamefont
  {Davison}, \citenamefont {Schalm},\ and\ \citenamefont
  {Zaanen}}]{Davison:2013txa}%
  \BibitemOpen
  \bibfield  {author} {\bibinfo {author} {\bibfnamefont {R.~A.}\ \bibnamefont
  {Davison}}, \bibinfo {author} {\bibfnamefont {K.}~\bibnamefont {Schalm}}, \
  and\ \bibinfo {author} {\bibfnamefont {J.}~\bibnamefont {Zaanen}},\ }\href
  {\doibase 10.1103/PhysRevB.89.245116} {\bibfield  {journal} {\bibinfo
  {journal} {Phys. Rev. B}\ }\textbf {\bibinfo {volume} {89}},\ \bibinfo
  {pages} {245116} (\bibinfo {year} {2014})},\ \Eprint
  {http://arxiv.org/abs/1311.2451} {arXiv:1311.2451 [hep-th]} \BibitemShut
  {NoStop}%
\bibitem [{\citenamefont {Jeong}\ \emph {et~al.}(2018)\citenamefont {Jeong},
  \citenamefont {Kim},\ and\ \citenamefont {Niu}}]{Jeong:2018tua}%
  \BibitemOpen
  \bibfield  {author} {\bibinfo {author} {\bibfnamefont {H.-S.}\ \bibnamefont
  {Jeong}}, \bibinfo {author} {\bibfnamefont {K.-Y.}\ \bibnamefont {Kim}}, \
  and\ \bibinfo {author} {\bibfnamefont {C.}~\bibnamefont {Niu}},\ }\href
  {\doibase 10.1007/JHEP10(2018)191} {\bibfield  {journal} {\bibinfo  {journal}
  {JHEP}\ }\textbf {\bibinfo {volume} {10}},\ \bibinfo {pages} {191} (\bibinfo
  {year} {2018})},\ \Eprint {http://arxiv.org/abs/1806.07739} {arXiv:1806.07739
  [hep-th]} \BibitemShut {NoStop}%
\bibitem [{\citenamefont {Ahn}\ \emph {et~al.}(2020)\citenamefont {Ahn},
  \citenamefont {Jeong}, \citenamefont {Ahn},\ and\ \citenamefont
  {Kim}}]{Ahn:2019lrh}%
  \BibitemOpen
  \bibfield  {author} {\bibinfo {author} {\bibfnamefont {Y.}~\bibnamefont
  {Ahn}}, \bibinfo {author} {\bibfnamefont {H.-S.}\ \bibnamefont {Jeong}},
  \bibinfo {author} {\bibfnamefont {D.}~\bibnamefont {Ahn}}, \ and\ \bibinfo
  {author} {\bibfnamefont {K.-Y.}\ \bibnamefont {Kim}},\ }\href {\doibase
  10.1007/JHEP04(2020)153} {\bibfield  {journal} {\bibinfo  {journal} {JHEP}\
  }\textbf {\bibinfo {volume} {04}},\ \bibinfo {pages} {153} (\bibinfo {year}
  {2020})},\ \Eprint {http://arxiv.org/abs/1907.12168} {arXiv:1907.12168
  [hep-th]} \BibitemShut {NoStop}%
\bibitem [{\citenamefont {Balm}\ \emph {et~al.}(2022)\citenamefont {Balm} \emph
  {et~al.}}]{Balm:2022bju}%
  \BibitemOpen
  \bibfield  {author} {\bibinfo {author} {\bibfnamefont {F.}~\bibnamefont
  {Balm}} \emph {et~al.},\ }\href@noop {} {\  (\bibinfo {year} {2022})},\
  \Eprint {http://arxiv.org/abs/2211.05492} {arXiv:2211.05492
  [cond-mat.str-el]} \BibitemShut {NoStop}%
\bibitem [{\citenamefont {Samanta}\ \emph {et~al.}(2022)\citenamefont
  {Samanta}, \citenamefont {Swain}, \citenamefont {Dou\c{c}ot}, \citenamefont
  {Policastro},\ and\ \citenamefont {Mukhopadhyay}}]{Samanta:2022rkx}%
  \BibitemOpen
  \bibfield  {author} {\bibinfo {author} {\bibfnamefont {S.}~\bibnamefont
  {Samanta}}, \bibinfo {author} {\bibfnamefont {H.}~\bibnamefont {Swain}},
  \bibinfo {author} {\bibfnamefont {B.}~\bibnamefont {Dou\c{c}ot}}, \bibinfo
  {author} {\bibfnamefont {G.}~\bibnamefont {Policastro}}, \ and\ \bibinfo
  {author} {\bibfnamefont {A.}~\bibnamefont {Mukhopadhyay}},\ }\href@noop {} {\
   (\bibinfo {year} {2022})},\ \Eprint {http://arxiv.org/abs/2206.01215}
  {arXiv:2206.01215 [cond-mat.str-el]} \BibitemShut {NoStop}%
\bibitem [{\citenamefont {Gubser}\ and\ \citenamefont
  {Rocha}(2010)}]{PhysRevD.81.046001}%
  \BibitemOpen
  \bibfield  {author} {\bibinfo {author} {\bibfnamefont {S.~S.}\ \bibnamefont
  {Gubser}}\ and\ \bibinfo {author} {\bibfnamefont {F.~D.}\ \bibnamefont
  {Rocha}},\ }\href {\doibase 10.1103/PhysRevD.81.046001} {\bibfield  {journal}
  {\bibinfo  {journal} {Phys. Rev. D}\ }\textbf {\bibinfo {volume} {81}},\
  \bibinfo {pages} {046001} (\bibinfo {year} {2010})}\BibitemShut {NoStop}%
\bibitem [{\citenamefont {Charmousis}\ \emph {et~al.}(2010)\citenamefont
  {Charmousis}, \citenamefont {Gouteraux}, \citenamefont {Kim}, \citenamefont
  {Kiritsis},\ and\ \citenamefont {Meyer}}]{Charmousis:2010zz}%
  \BibitemOpen
  \bibfield  {author} {\bibinfo {author} {\bibfnamefont {C.}~\bibnamefont
  {Charmousis}}, \bibinfo {author} {\bibfnamefont {B.}~\bibnamefont
  {Gouteraux}}, \bibinfo {author} {\bibfnamefont {B.~S.}\ \bibnamefont {Kim}},
  \bibinfo {author} {\bibfnamefont {E.}~\bibnamefont {Kiritsis}}, \ and\
  \bibinfo {author} {\bibfnamefont {R.}~\bibnamefont {Meyer}},\ }\href
  {\doibase 10.1007/JHEP11(2010)151} {\bibfield  {journal} {\bibinfo  {journal}
  {JHEP}\ }\textbf {\bibinfo {volume} {11}},\ \bibinfo {pages} {151} (\bibinfo
  {year} {2010})},\ \Eprint {http://arxiv.org/abs/1005.4690} {arXiv:1005.4690
  [hep-th]} \BibitemShut {NoStop}%
\bibitem [{\citenamefont {Gout\'eraux}(2014)}]{Gouteraux:2014hca}%
  \BibitemOpen
  \bibfield  {author} {\bibinfo {author} {\bibfnamefont {B.}~\bibnamefont
  {Gout\'eraux}},\ }\href {\doibase 10.1007/JHEP04(2014)181} {\bibfield
  {journal} {\bibinfo  {journal} {JHEP}\ }\textbf {\bibinfo {volume} {04}},\
  \bibinfo {pages} {181} (\bibinfo {year} {2014})},\ \Eprint
  {http://arxiv.org/abs/1401.5436} {arXiv:1401.5436 [hep-th]} \BibitemShut
  {NoStop}%
\bibitem [{\citenamefont {Iqbal}\ \emph {et~al.}(2012)\citenamefont {Iqbal},
  \citenamefont {Liu},\ and\ \citenamefont {Mezei}}]{Iqbal:2011in}%
  \BibitemOpen
  \bibfield  {author} {\bibinfo {author} {\bibfnamefont {N.}~\bibnamefont
  {Iqbal}}, \bibinfo {author} {\bibfnamefont {H.}~\bibnamefont {Liu}}, \ and\
  \bibinfo {author} {\bibfnamefont {M.}~\bibnamefont {Mezei}},\ }\href
  {\doibase 10.1007/JHEP04(2012)086} {\bibfield  {journal} {\bibinfo  {journal}
  {JHEP}\ }\textbf {\bibinfo {volume} {04}},\ \bibinfo {pages} {086} (\bibinfo
  {year} {2012})},\ \Eprint {http://arxiv.org/abs/1105.4621} {arXiv:1105.4621
  [hep-th]} \BibitemShut {NoStop}%
\bibitem [{\citenamefont {Faulkner}\ \emph {et~al.}(2011)\citenamefont
  {Faulkner}, \citenamefont {Liu}, \citenamefont {McGreevy},\ and\
  \citenamefont {Vegh}}]{PhysRevD.83.125002}%
  \BibitemOpen
  \bibfield  {author} {\bibinfo {author} {\bibfnamefont {T.}~\bibnamefont
  {Faulkner}}, \bibinfo {author} {\bibfnamefont {H.}~\bibnamefont {Liu}},
  \bibinfo {author} {\bibfnamefont {J.}~\bibnamefont {McGreevy}}, \ and\
  \bibinfo {author} {\bibfnamefont {D.}~\bibnamefont {Vegh}},\ }\href {\doibase
  10.1103/PhysRevD.83.125002} {\bibfield  {journal} {\bibinfo  {journal} {Phys.
  Rev. D}\ }\textbf {\bibinfo {volume} {83}},\ \bibinfo {pages} {125002}
  (\bibinfo {year} {2011})}\BibitemShut {NoStop}%
\bibitem [{\citenamefont {Chowdhury}\ \emph {et~al.}(2022)\citenamefont
  {Chowdhury}, \citenamefont {Georges}, \citenamefont {Parcollet},\ and\
  \citenamefont {Sachdev}}]{RevModPhys.94.035004}%
  \BibitemOpen
  \bibfield  {author} {\bibinfo {author} {\bibfnamefont {D.}~\bibnamefont
  {Chowdhury}}, \bibinfo {author} {\bibfnamefont {A.}~\bibnamefont {Georges}},
  \bibinfo {author} {\bibfnamefont {O.}~\bibnamefont {Parcollet}}, \ and\
  \bibinfo {author} {\bibfnamefont {S.}~\bibnamefont {Sachdev}},\ }\href
  {\doibase 10.1103/RevModPhys.94.035004} {\bibfield  {journal} {\bibinfo
  {journal} {Rev. Mod. Phys.}\ }\textbf {\bibinfo {volume} {94}},\ \bibinfo
  {pages} {035004} (\bibinfo {year} {2022})}\BibitemShut {NoStop}%
\bibitem [{\citenamefont {Anderson}(2013)}]{10.1063/PT.3.1929}%
  \BibitemOpen
  \bibfield  {author} {\bibinfo {author} {\bibfnamefont {P.~W.}\ \bibnamefont
  {Anderson}},\ }\href {\doibase 10.1063/PT.3.1929} {\bibfield  {journal}
  {\bibinfo  {journal} {Physics Today}\ }\textbf {\bibinfo {volume} {66}},\
  \bibinfo {pages} {9} (\bibinfo {year} {2013})}\BibitemShut {NoStop}%
\bibitem [{\citenamefont {Khveshchenko}(2021)}]{Khveshchenko:2020pma}%
  \BibitemOpen
  \bibfield  {author} {\bibinfo {author} {\bibfnamefont {D.~V.}\ \bibnamefont
  {Khveshchenko}},\ }\href@noop {} {\bibfield  {journal} {\bibinfo  {journal}
  {Lith. J. Phys.}\ }\textbf {\bibinfo {volume} {61}},\ \bibinfo {pages} {1}
  (\bibinfo {year} {2021})},\ \Eprint {http://arxiv.org/abs/2011.11617}
  {arXiv:2011.11617 [cond-mat.str-el]} \BibitemShut {NoStop}%
\bibitem [{\citenamefont {Khveshchenko}(2016)}]{Khveshchenko:2016kmo}%
  \BibitemOpen
  \bibfield  {author} {\bibinfo {author} {\bibfnamefont {D.~V.}\ \bibnamefont
  {Khveshchenko}},\ }\href {\doibase 10.3952/physics.v56i3.3363} {\bibfield
  {journal} {\bibinfo  {journal} {Lith. J. Phys.}\ }\textbf {\bibinfo {volume}
  {56}},\ \bibinfo {pages} {125} (\bibinfo {year} {2016})},\ \Eprint
  {http://arxiv.org/abs/1603.09741} {arXiv:1603.09741 [cond-mat.str-el]}
  \BibitemShut {NoStop}%
\bibitem [{\citenamefont
  {Khveshchenko}(2015{\natexlab{a}})}]{Khveshchenko:2015xea}%
  \BibitemOpen
  \bibfield  {author} {\bibinfo {author} {\bibfnamefont {D.~V.}\ \bibnamefont
  {Khveshchenko}},\ }\href {\doibase 10.1209/0295-5075/111/17003} {\bibfield
  {journal} {\bibinfo  {journal} {EPL}\ }\textbf {\bibinfo {volume} {111}},\
  \bibinfo {pages} {1700} (\bibinfo {year} {2015}{\natexlab{a}})},\ \Eprint
  {http://arxiv.org/abs/1502.03375} {arXiv:1502.03375 [cond-mat.str-el]}
  \BibitemShut {NoStop}%
\bibitem [{\citenamefont
  {Khveshchenko}(2015{\natexlab{b}})}]{Khveshchenko:2014nka}%
  \BibitemOpen
  \bibfield  {author} {\bibinfo {author} {\bibfnamefont {D.~V.}\ \bibnamefont
  {Khveshchenko}},\ }\href {\doibase 10.3952/physics.v55i3.3150} {\bibfield
  {journal} {\bibinfo  {journal} {Lith. J. Phys.}\ }\textbf {\bibinfo {volume}
  {55}},\ \bibinfo {pages} {208} (\bibinfo {year} {2015}{\natexlab{b}})},\
  \Eprint {http://arxiv.org/abs/1404.7000} {arXiv:1404.7000 [cond-mat.str-el]}
  \BibitemShut {NoStop}%
\bibitem [{\citenamefont {Khveshchenko}(2012)}]{Khveshchenko:2012yt}%
  \BibitemOpen
  \bibfield  {author} {\bibinfo {author} {\bibfnamefont {D.~V.}\ \bibnamefont
  {Khveshchenko}},\ }\href {\doibase 10.1103/PhysRevB.86.115115} {\bibfield
  {journal} {\bibinfo  {journal} {Phys. Rev. B}\ }\textbf {\bibinfo {volume}
  {86}},\ \bibinfo {pages} {115115} (\bibinfo {year} {2012})},\ \Eprint
  {http://arxiv.org/abs/1205.4420} {arXiv:1205.4420 [cond-mat.str-el]}
  \BibitemShut {NoStop}%
\bibitem [{\citenamefont {Phillips}\ \emph {et~al.}(2022)\citenamefont
  {Phillips}, \citenamefont {Hussey},\ and\ \citenamefont
  {Abbamonte}}]{Phillips:2022nxs}%
  \BibitemOpen
  \bibfield  {author} {\bibinfo {author} {\bibfnamefont {P.~W.}\ \bibnamefont
  {Phillips}}, \bibinfo {author} {\bibfnamefont {N.~E.}\ \bibnamefont
  {Hussey}}, \ and\ \bibinfo {author} {\bibfnamefont {P.}~\bibnamefont
  {Abbamonte}},\ }\href {\doibase 10.1126/science.abh4273} {\bibfield
  {journal} {\bibinfo  {journal} {Science}\ }\textbf {\bibinfo {volume}
  {377}},\ \bibinfo {pages} {abh4273} (\bibinfo {year} {2022})},\ \Eprint
  {http://arxiv.org/abs/2205.12979} {arXiv:2205.12979 [cond-mat.str-el]}
  \BibitemShut {NoStop}%
\bibitem [{\citenamefont {Hartnoll}\ and\ \citenamefont
  {Mackenzie}(2022)}]{Hartnoll:2021ydi}%
  \BibitemOpen
  \bibfield  {author} {\bibinfo {author} {\bibfnamefont {S.~A.}\ \bibnamefont
  {Hartnoll}}\ and\ \bibinfo {author} {\bibfnamefont {A.~P.}\ \bibnamefont
  {Mackenzie}},\ }\href {\doibase 10.1103/RevModPhys.94.041002} {\bibfield
  {journal} {\bibinfo  {journal} {Rev. Mod. Phys.}\ }\textbf {\bibinfo {volume}
  {94}},\ \bibinfo {pages} {041002} (\bibinfo {year} {2022})},\ \Eprint
  {http://arxiv.org/abs/2107.07802} {arXiv:2107.07802 [cond-mat.str-el]}
  \BibitemShut {NoStop}%
\bibitem [{\citenamefont {Chien}\ \emph {et~al.}(1991)\citenamefont {Chien},
  \citenamefont {Wang},\ and\ \citenamefont {Ong}}]{PhysRevLett.67.2088}%
  \BibitemOpen
  \bibfield  {author} {\bibinfo {author} {\bibfnamefont {T.~R.}\ \bibnamefont
  {Chien}}, \bibinfo {author} {\bibfnamefont {Z.~Z.}\ \bibnamefont {Wang}}, \
  and\ \bibinfo {author} {\bibfnamefont {N.~P.}\ \bibnamefont {Ong}},\ }\href
  {\doibase 10.1103/PhysRevLett.67.2088} {\bibfield  {journal} {\bibinfo
  {journal} {Phys. Rev. Lett.}\ }\textbf {\bibinfo {volume} {67}},\ \bibinfo
  {pages} {2088} (\bibinfo {year} {1991})}\BibitemShut {NoStop}%
\bibitem [{\citenamefont {Harris}\ \emph {et~al.}(1995)\citenamefont {Harris},
  \citenamefont {Yan}, \citenamefont {Matl}, \citenamefont {Ong}, \citenamefont
  {Anderson}, \citenamefont {Kimura},\ and\ \citenamefont
  {Kitazawa}}]{PhysRevLett.75.1391}%
  \BibitemOpen
  \bibfield  {author} {\bibinfo {author} {\bibfnamefont {J.~M.}\ \bibnamefont
  {Harris}}, \bibinfo {author} {\bibfnamefont {Y.~F.}\ \bibnamefont {Yan}},
  \bibinfo {author} {\bibfnamefont {P.}~\bibnamefont {Matl}}, \bibinfo {author}
  {\bibfnamefont {N.~P.}\ \bibnamefont {Ong}}, \bibinfo {author} {\bibfnamefont
  {P.~W.}\ \bibnamefont {Anderson}}, \bibinfo {author} {\bibfnamefont
  {T.}~\bibnamefont {Kimura}}, \ and\ \bibinfo {author} {\bibfnamefont
  {K.}~\bibnamefont {Kitazawa}},\ }\href {\doibase 10.1103/PhysRevLett.75.1391}
  {\bibfield  {journal} {\bibinfo  {journal} {Phys. Rev. Lett.}\ }\textbf
  {\bibinfo {volume} {75}},\ \bibinfo {pages} {1391} (\bibinfo {year}
  {1995})}\BibitemShut {NoStop}%
\bibitem [{\citenamefont {Anderson}(1991)}]{PhysRevLett.67.2092}%
  \BibitemOpen
  \bibfield  {author} {\bibinfo {author} {\bibfnamefont {P.~W.}\ \bibnamefont
  {Anderson}},\ }\href {\doibase 10.1103/PhysRevLett.67.2092} {\bibfield
  {journal} {\bibinfo  {journal} {Phys. Rev. Lett.}\ }\textbf {\bibinfo
  {volume} {67}},\ \bibinfo {pages} {2092} (\bibinfo {year}
  {1991})}\BibitemShut {NoStop}%
\bibitem [{\citenamefont {Blake}\ and\ \citenamefont
  {Donos}(2015)}]{Blake:2014yla}%
  \BibitemOpen
  \bibfield  {author} {\bibinfo {author} {\bibfnamefont {M.}~\bibnamefont
  {Blake}}\ and\ \bibinfo {author} {\bibfnamefont {A.}~\bibnamefont {Donos}},\
  }\href {\doibase 10.1103/PhysRevLett.114.021601} {\bibfield  {journal}
  {\bibinfo  {journal} {Phys. Rev. Lett.}\ }\textbf {\bibinfo {volume} {114}},\
  \bibinfo {pages} {021601} (\bibinfo {year} {2015})},\ \Eprint
  {http://arxiv.org/abs/1406.1659} {arXiv:1406.1659 [hep-th]} \BibitemShut
  {NoStop}%
\bibitem [{\citenamefont {Donos}\ and\ \citenamefont
  {Gauntlett}(2014)}]{Donos:2014cya}%
  \BibitemOpen
  \bibfield  {author} {\bibinfo {author} {\bibfnamefont {A.}~\bibnamefont
  {Donos}}\ and\ \bibinfo {author} {\bibfnamefont {J.~P.}\ \bibnamefont
  {Gauntlett}},\ }\href {\doibase 10.1007/JHEP11(2014)081} {\bibfield
  {journal} {\bibinfo  {journal} {JHEP}\ }\textbf {\bibinfo {volume} {11}},\
  \bibinfo {pages} {081} (\bibinfo {year} {2014})},\ \Eprint
  {http://arxiv.org/abs/1406.4742} {arXiv:1406.4742 [hep-th]} \BibitemShut
  {NoStop}%
\bibitem [{\citenamefont {Baggioli}\ and\ \citenamefont
  {Pujolas}(2016)}]{Baggioli:2016oju}%
  \BibitemOpen
  \bibfield  {author} {\bibinfo {author} {\bibfnamefont {M.}~\bibnamefont
  {Baggioli}}\ and\ \bibinfo {author} {\bibfnamefont {O.}~\bibnamefont
  {Pujolas}},\ }\href {\doibase 10.1007/JHEP12(2016)107} {\bibfield  {journal}
  {\bibinfo  {journal} {JHEP}\ }\textbf {\bibinfo {volume} {12}},\ \bibinfo
  {pages} {107} (\bibinfo {year} {2016})},\ \Eprint
  {http://arxiv.org/abs/1604.08915} {arXiv:1604.08915 [hep-th]} \BibitemShut
  {NoStop}%
\bibitem [{\citenamefont {Blake}(2015)}]{Blake:2015epa}%
  \BibitemOpen
  \bibfield  {author} {\bibinfo {author} {\bibfnamefont {M.}~\bibnamefont
  {Blake}},\ }\href {\doibase 10.1007/JHEP09(2015)010} {\bibfield  {journal}
  {\bibinfo  {journal} {JHEP}\ }\textbf {\bibinfo {volume} {09}},\ \bibinfo
  {pages} {010} (\bibinfo {year} {2015})},\ \Eprint
  {http://arxiv.org/abs/1505.06992} {arXiv:1505.06992 [hep-th]} \BibitemShut
  {NoStop}%
\bibitem [{\citenamefont {Davison}\ and\ \citenamefont
  {Gout\'eraux}(2015)}]{Davison:2015bea}%
  \BibitemOpen
  \bibfield  {author} {\bibinfo {author} {\bibfnamefont {R.~A.}\ \bibnamefont
  {Davison}}\ and\ \bibinfo {author} {\bibfnamefont {B.}~\bibnamefont
  {Gout\'eraux}},\ }\href {\doibase 10.1007/JHEP09(2015)090} {\bibfield
  {journal} {\bibinfo  {journal} {JHEP}\ }\textbf {\bibinfo {volume} {09}},\
  \bibinfo {pages} {090} (\bibinfo {year} {2015})},\ \Eprint
  {http://arxiv.org/abs/1505.05092} {arXiv:1505.05092 [hep-th]} \BibitemShut
  {NoStop}%
\bibitem [{\citenamefont {Marel}\ \emph {et~al.}(2003)\citenamefont {Marel},
  \citenamefont {Molegraaf}, \citenamefont {Zaanen}, \citenamefont {Nussinov},
  \citenamefont {Carbone}, \citenamefont {Damascelli}, \citenamefont {Eisaki},
  \citenamefont {Greven}, \citenamefont {Kes},\ and\ \citenamefont
  {Li}}]{Marel2003}%
  \BibitemOpen
  \bibfield  {author} {\bibinfo {author} {\bibfnamefont {D.~v.~d.}\
  \bibnamefont {Marel}}, \bibinfo {author} {\bibfnamefont {H.~J.~A.}\
  \bibnamefont {Molegraaf}}, \bibinfo {author} {\bibfnamefont {J.}~\bibnamefont
  {Zaanen}}, \bibinfo {author} {\bibfnamefont {Z.}~\bibnamefont {Nussinov}},
  \bibinfo {author} {\bibfnamefont {F.}~\bibnamefont {Carbone}}, \bibinfo
  {author} {\bibfnamefont {A.}~\bibnamefont {Damascelli}}, \bibinfo {author}
  {\bibfnamefont {H.}~\bibnamefont {Eisaki}}, \bibinfo {author} {\bibfnamefont
  {M.}~\bibnamefont {Greven}}, \bibinfo {author} {\bibfnamefont {P.~H.}\
  \bibnamefont {Kes}}, \ and\ \bibinfo {author} {\bibfnamefont
  {M.}~\bibnamefont {Li}},\ }\href {\doibase 10.1038/nature01978} {\bibfield
  {journal} {\bibinfo  {journal} {Nature}\ }\textbf {\bibinfo {volume} {425}},\
  \bibinfo {pages} {271} (\bibinfo {year} {2003})}\BibitemShut {NoStop}%
\bibitem [{\citenamefont {Hartnoll}\ and\ \citenamefont
  {Karch}(2015)}]{Hartnoll:2015sea}%
  \BibitemOpen
  \bibfield  {author} {\bibinfo {author} {\bibfnamefont {S.~A.}\ \bibnamefont
  {Hartnoll}}\ and\ \bibinfo {author} {\bibfnamefont {A.}~\bibnamefont
  {Karch}},\ }\href {\doibase 10.1103/PhysRevB.91.155126} {\bibfield  {journal}
  {\bibinfo  {journal} {Phys. Rev. B}\ }\textbf {\bibinfo {volume} {91}},\
  \bibinfo {pages} {155126} (\bibinfo {year} {2015})},\ \Eprint
  {http://arxiv.org/abs/1501.03165} {arXiv:1501.03165 [cond-mat.str-el]}
  \BibitemShut {NoStop}%
\bibitem [{\citenamefont {Karch}\ \emph {et~al.}(2016)\citenamefont {Karch},
  \citenamefont {Limtragool},\ and\ \citenamefont {Phillips}}]{Karch:2015zqd}%
  \BibitemOpen
  \bibfield  {author} {\bibinfo {author} {\bibfnamefont {A.}~\bibnamefont
  {Karch}}, \bibinfo {author} {\bibfnamefont {K.}~\bibnamefont {Limtragool}}, \
  and\ \bibinfo {author} {\bibfnamefont {P.~W.}\ \bibnamefont {Phillips}},\
  }\href {\doibase 10.1007/JHEP03(2016)175} {\bibfield  {journal} {\bibinfo
  {journal} {JHEP}\ }\textbf {\bibinfo {volume} {03}},\ \bibinfo {pages} {175}
  (\bibinfo {year} {2016})},\ \Eprint {http://arxiv.org/abs/1511.02868}
  {arXiv:1511.02868 [cond-mat.str-el]} \BibitemShut {NoStop}%
\bibitem [{\citenamefont {Hartnoll}(2015)}]{Hartnoll2015}%
  \BibitemOpen
  \bibfield  {author} {\bibinfo {author} {\bibfnamefont {S.~A.}\ \bibnamefont
  {Hartnoll}},\ }\href {\doibase 10.1038/nphys3174} {\bibfield  {journal}
  {\bibinfo  {journal} {Nature Physics}\ }\textbf {\bibinfo {volume} {11}},\
  \bibinfo {pages} {54} (\bibinfo {year} {2015})}\BibitemShut {NoStop}%
\bibitem [{\citenamefont {Baggioli}\ \emph {et~al.}(2017)\citenamefont
  {Baggioli}, \citenamefont {Gout\'eraux}, \citenamefont {Kiritsis},\ and\
  \citenamefont {Li}}]{Baggioli:2016pia}%
  \BibitemOpen
  \bibfield  {author} {\bibinfo {author} {\bibfnamefont {M.}~\bibnamefont
  {Baggioli}}, \bibinfo {author} {\bibfnamefont {B.}~\bibnamefont
  {Gout\'eraux}}, \bibinfo {author} {\bibfnamefont {E.}~\bibnamefont
  {Kiritsis}}, \ and\ \bibinfo {author} {\bibfnamefont {W.-J.}\ \bibnamefont
  {Li}},\ }\href {\doibase 10.1007/JHEP03(2017)170} {\bibfield  {journal}
  {\bibinfo  {journal} {JHEP}\ }\textbf {\bibinfo {volume} {03}},\ \bibinfo
  {pages} {170} (\bibinfo {year} {2017})},\ \Eprint
  {http://arxiv.org/abs/1612.05500} {arXiv:1612.05500 [hep-th]} \BibitemShut
  {NoStop}%
\bibitem [{\citenamefont {Amoretti}\ \emph {et~al.}(2018)\citenamefont
  {Amoretti}, \citenamefont {Are\'an}, \citenamefont {Gout\'eraux},\ and\
  \citenamefont {Musso}}]{PhysRevLett.120.171603}%
  \BibitemOpen
  \bibfield  {author} {\bibinfo {author} {\bibfnamefont {A.}~\bibnamefont
  {Amoretti}}, \bibinfo {author} {\bibfnamefont {D.}~\bibnamefont {Are\'an}},
  \bibinfo {author} {\bibfnamefont {B.}~\bibnamefont {Gout\'eraux}}, \ and\
  \bibinfo {author} {\bibfnamefont {D.}~\bibnamefont {Musso}},\ }\href
  {\doibase 10.1103/PhysRevLett.120.171603} {\bibfield  {journal} {\bibinfo
  {journal} {Phys. Rev. Lett.}\ }\textbf {\bibinfo {volume} {120}},\ \bibinfo
  {pages} {171603} (\bibinfo {year} {2018})}\BibitemShut {NoStop}%
\bibitem [{\citenamefont {van Heumen}\ \emph {et~al.}(2022)\citenamefont {van
  Heumen}, \citenamefont {Feng}, \citenamefont {Cassanelli}, \citenamefont
  {Neubrand}, \citenamefont {de~Jager}, \citenamefont {Berben}, \citenamefont
  {Huang}, \citenamefont {Kondo}, \citenamefont {Takeuchi},\ and\ \citenamefont
  {Zaanen}}]{PhysRevB.106.054515}%
  \BibitemOpen
  \bibfield  {author} {\bibinfo {author} {\bibfnamefont {E.}~\bibnamefont {van
  Heumen}}, \bibinfo {author} {\bibfnamefont {X.}~\bibnamefont {Feng}},
  \bibinfo {author} {\bibfnamefont {S.}~\bibnamefont {Cassanelli}}, \bibinfo
  {author} {\bibfnamefont {L.}~\bibnamefont {Neubrand}}, \bibinfo {author}
  {\bibfnamefont {L.}~\bibnamefont {de~Jager}}, \bibinfo {author}
  {\bibfnamefont {M.}~\bibnamefont {Berben}}, \bibinfo {author} {\bibfnamefont
  {Y.}~\bibnamefont {Huang}}, \bibinfo {author} {\bibfnamefont
  {T.}~\bibnamefont {Kondo}}, \bibinfo {author} {\bibfnamefont
  {T.}~\bibnamefont {Takeuchi}}, \ and\ \bibinfo {author} {\bibfnamefont
  {J.}~\bibnamefont {Zaanen}},\ }\href {\doibase 10.1103/PhysRevB.106.054515}
  {\bibfield  {journal} {\bibinfo  {journal} {Phys. Rev. B}\ }\textbf {\bibinfo
  {volume} {106}},\ \bibinfo {pages} {054515} (\bibinfo {year}
  {2022})}\BibitemShut {NoStop}%
\bibitem [{\citenamefont {Bruin}\ \emph {et~al.}(2013)\citenamefont {Bruin},
  \citenamefont {Sakai}, \citenamefont {Perry},\ and\ \citenamefont
  {Mackenzie}}]{doi:10.1126/science.1227612}%
  \BibitemOpen
  \bibfield  {author} {\bibinfo {author} {\bibfnamefont {J.~A.~N.}\
  \bibnamefont {Bruin}}, \bibinfo {author} {\bibfnamefont {H.}~\bibnamefont
  {Sakai}}, \bibinfo {author} {\bibfnamefont {R.~S.}\ \bibnamefont {Perry}}, \
  and\ \bibinfo {author} {\bibfnamefont {A.~P.}\ \bibnamefont {Mackenzie}},\
  }\href {\doibase 10.1126/science.1227612} {\bibfield  {journal} {\bibinfo
  {journal} {Science}\ }\textbf {\bibinfo {volume} {339}},\ \bibinfo {pages}
  {804} (\bibinfo {year} {2013})}\BibitemShut {NoStop}%
\bibitem [{\citenamefont {Rullier-Albenque}\ \emph {et~al.}(2003)\citenamefont
  {Rullier-Albenque}, \citenamefont {Alloul},\ and\ \citenamefont
  {Tourbot}}]{PhysRevLett.91.047001}%
  \BibitemOpen
  \bibfield  {author} {\bibinfo {author} {\bibfnamefont {F.}~\bibnamefont
  {Rullier-Albenque}}, \bibinfo {author} {\bibfnamefont {H.}~\bibnamefont
  {Alloul}}, \ and\ \bibinfo {author} {\bibfnamefont {R.}~\bibnamefont
  {Tourbot}},\ }\href {\doibase 10.1103/PhysRevLett.91.047001} {\bibfield
  {journal} {\bibinfo  {journal} {Phys. Rev. Lett.}\ }\textbf {\bibinfo
  {volume} {91}},\ \bibinfo {pages} {047001} (\bibinfo {year}
  {2003})}\BibitemShut {NoStop}%
\bibitem [{\citenamefont {Delacr\'etaz}\ \emph {et~al.}(2022)\citenamefont
  {Delacr\'etaz}, \citenamefont {Gout\'eraux},\ and\ \citenamefont
  {Ziogas}}]{PhysRevLett.128.141601}%
  \BibitemOpen
  \bibfield  {author} {\bibinfo {author} {\bibfnamefont {L.~V.}\ \bibnamefont
  {Delacr\'etaz}}, \bibinfo {author} {\bibfnamefont {B.}~\bibnamefont
  {Gout\'eraux}}, \ and\ \bibinfo {author} {\bibfnamefont {V.}~\bibnamefont
  {Ziogas}},\ }\href {\doibase 10.1103/PhysRevLett.128.141601} {\bibfield
  {journal} {\bibinfo  {journal} {Phys. Rev. Lett.}\ }\textbf {\bibinfo
  {volume} {128}},\ \bibinfo {pages} {141601} (\bibinfo {year}
  {2022})}\BibitemShut {NoStop}%
\bibitem [{\citenamefont {Baggioli}\ and\ \citenamefont
  {Gout\'eraux}(2023)}]{RevModPhys.95.011001}%
  \BibitemOpen
  \bibfield  {author} {\bibinfo {author} {\bibfnamefont {M.}~\bibnamefont
  {Baggioli}}\ and\ \bibinfo {author} {\bibfnamefont {B.}~\bibnamefont
  {Gout\'eraux}},\ }\href {\doibase 10.1103/RevModPhys.95.011001} {\bibfield
  {journal} {\bibinfo  {journal} {Rev. Mod. Phys.}\ }\textbf {\bibinfo {volume}
  {95}},\ \bibinfo {pages} {011001} (\bibinfo {year} {2023})}\BibitemShut
  {NoStop}%
\bibitem [{\citenamefont {Amoretti}\ \emph
  {et~al.}(2020{\natexlab{a}})\citenamefont {Amoretti}, \citenamefont
  {Meinero}, \citenamefont {Brattan}, \citenamefont {Caglieris}, \citenamefont
  {Giannini}, \citenamefont {Affronte}, \citenamefont {Hess}, \citenamefont
  {Buechner}, \citenamefont {Magnoli},\ and\ \citenamefont
  {Putti}}]{Amoretti:2019buu}%
  \BibitemOpen
  \bibfield  {author} {\bibinfo {author} {\bibfnamefont {A.}~\bibnamefont
  {Amoretti}}, \bibinfo {author} {\bibfnamefont {M.}~\bibnamefont {Meinero}},
  \bibinfo {author} {\bibfnamefont {D.~K.}\ \bibnamefont {Brattan}}, \bibinfo
  {author} {\bibfnamefont {F.}~\bibnamefont {Caglieris}}, \bibinfo {author}
  {\bibfnamefont {E.}~\bibnamefont {Giannini}}, \bibinfo {author}
  {\bibfnamefont {M.}~\bibnamefont {Affronte}}, \bibinfo {author}
  {\bibfnamefont {C.}~\bibnamefont {Hess}}, \bibinfo {author} {\bibfnamefont
  {B.}~\bibnamefont {Buechner}}, \bibinfo {author} {\bibfnamefont
  {N.}~\bibnamefont {Magnoli}}, \ and\ \bibinfo {author} {\bibfnamefont
  {M.}~\bibnamefont {Putti}},\ }\href {\doibase
  10.1103/PhysRevResearch.2.023387} {\bibfield  {journal} {\bibinfo  {journal}
  {Phys. Rev. Res.}\ }\textbf {\bibinfo {volume} {2}},\ \bibinfo {pages}
  {023387} (\bibinfo {year} {2020}{\natexlab{a}})},\ \Eprint
  {http://arxiv.org/abs/1909.07991} {arXiv:1909.07991 [cond-mat.str-el]}
  \BibitemShut {NoStop}%
\bibitem [{\citenamefont {Anantua}\ \emph {et~al.}(2013)\citenamefont
  {Anantua}, \citenamefont {Hartnoll}, \citenamefont {Martin},\ and\
  \citenamefont {Ramirez}}]{Anantua:2012nj}%
  \BibitemOpen
  \bibfield  {author} {\bibinfo {author} {\bibfnamefont {R.~J.}\ \bibnamefont
  {Anantua}}, \bibinfo {author} {\bibfnamefont {S.~A.}\ \bibnamefont
  {Hartnoll}}, \bibinfo {author} {\bibfnamefont {V.~L.}\ \bibnamefont
  {Martin}}, \ and\ \bibinfo {author} {\bibfnamefont {D.~M.}\ \bibnamefont
  {Ramirez}},\ }\href {\doibase 10.1007/JHEP03(2013)104} {\bibfield  {journal}
  {\bibinfo  {journal} {JHEP}\ }\textbf {\bibinfo {volume} {03}},\ \bibinfo
  {pages} {104} (\bibinfo {year} {2013})},\ \Eprint
  {http://arxiv.org/abs/1210.1590} {arXiv:1210.1590 [hep-th]} \BibitemShut
  {NoStop}%
\bibitem [{\citenamefont {Jeong}\ and\ \citenamefont
  {Kim}(2022)}]{Jeong:2021wiu}%
  \BibitemOpen
  \bibfield  {author} {\bibinfo {author} {\bibfnamefont {H.-S.}\ \bibnamefont
  {Jeong}}\ and\ \bibinfo {author} {\bibfnamefont {K.-Y.}\ \bibnamefont
  {Kim}},\ }\href {\doibase 10.1007/JHEP03(2022)060} {\bibfield  {journal}
  {\bibinfo  {journal} {JHEP}\ }\textbf {\bibinfo {volume} {03}},\ \bibinfo
  {pages} {060} (\bibinfo {year} {2022})},\ \Eprint
  {http://arxiv.org/abs/2112.01153} {arXiv:2112.01153 [hep-th]} \BibitemShut
  {NoStop}%
\bibitem [{\citenamefont {Wu}(2011)}]{Wu:2011cy}%
  \BibitemOpen
  \bibfield  {author} {\bibinfo {author} {\bibfnamefont {J.-P.}\ \bibnamefont
  {Wu}},\ }\href {\doibase 10.1103/PhysRevD.84.064008} {\bibfield  {journal}
  {\bibinfo  {journal} {Phys. Rev. D}\ }\textbf {\bibinfo {volume} {84}},\
  \bibinfo {pages} {064008} (\bibinfo {year} {2011})},\ \Eprint
  {http://arxiv.org/abs/1108.6134} {arXiv:1108.6134 [hep-th]} \BibitemShut
  {NoStop}%
\bibitem [{\citenamefont {Jeong}\ \emph {et~al.}(2020)\citenamefont {Jeong},
  \citenamefont {Kim}, \citenamefont {Seo}, \citenamefont {Sin},\ and\
  \citenamefont {Wu}}]{Jeong:2019zab}%
  \BibitemOpen
  \bibfield  {author} {\bibinfo {author} {\bibfnamefont {H.-S.}\ \bibnamefont
  {Jeong}}, \bibinfo {author} {\bibfnamefont {K.-Y.}\ \bibnamefont {Kim}},
  \bibinfo {author} {\bibfnamefont {Y.}~\bibnamefont {Seo}}, \bibinfo {author}
  {\bibfnamefont {S.-J.}\ \bibnamefont {Sin}}, \ and\ \bibinfo {author}
  {\bibfnamefont {S.-Y.}\ \bibnamefont {Wu}},\ }\href {\doibase
  10.1103/PhysRevD.102.026017} {\bibfield  {journal} {\bibinfo  {journal}
  {Phys. Rev. D}\ }\textbf {\bibinfo {volume} {102}},\ \bibinfo {pages}
  {026017} (\bibinfo {year} {2020})},\ \Eprint
  {http://arxiv.org/abs/1910.11034} {arXiv:1910.11034 [hep-th]} \BibitemShut
  {NoStop}%
\bibitem [{\citenamefont {Kiritsis}\ and\ \citenamefont
  {Li}(2017)}]{Kiritsis:2016cpm}%
  \BibitemOpen
  \bibfield  {author} {\bibinfo {author} {\bibfnamefont {E.}~\bibnamefont
  {Kiritsis}}\ and\ \bibinfo {author} {\bibfnamefont {L.}~\bibnamefont {Li}},\
  }\href {\doibase 10.1088/1751-8121/aa59c6} {\bibfield  {journal} {\bibinfo
  {journal} {J. Phys. A}\ }\textbf {\bibinfo {volume} {50}},\ \bibinfo {pages}
  {115402} (\bibinfo {year} {2017})},\ \Eprint
  {http://arxiv.org/abs/1608.02598} {arXiv:1608.02598 [cond-mat.str-el]}
  \BibitemShut {NoStop}%
\bibitem [{\citenamefont {Cremonini}\ \emph {et~al.}(2017)\citenamefont
  {Cremonini}, \citenamefont {Hoover},\ and\ \citenamefont
  {Li}}]{Cremonini:2017qwq}%
  \BibitemOpen
  \bibfield  {author} {\bibinfo {author} {\bibfnamefont {S.}~\bibnamefont
  {Cremonini}}, \bibinfo {author} {\bibfnamefont {A.}~\bibnamefont {Hoover}}, \
  and\ \bibinfo {author} {\bibfnamefont {L.}~\bibnamefont {Li}},\ }\href
  {\doibase 10.1007/JHEP10(2017)133} {\bibfield  {journal} {\bibinfo  {journal}
  {JHEP}\ }\textbf {\bibinfo {volume} {10}},\ \bibinfo {pages} {133} (\bibinfo
  {year} {2017})},\ \Eprint {http://arxiv.org/abs/1707.01505} {arXiv:1707.01505
  [hep-th]} \BibitemShut {NoStop}%
\bibitem [{\citenamefont {Baggioli}\ and\ \citenamefont
  {Pujol\`as}(2015)}]{PhysRevLett.114.251602}%
  \BibitemOpen
  \bibfield  {author} {\bibinfo {author} {\bibfnamefont {M.}~\bibnamefont
  {Baggioli}}\ and\ \bibinfo {author} {\bibfnamefont {O.}~\bibnamefont
  {Pujol\`as}},\ }\href {\doibase 10.1103/PhysRevLett.114.251602} {\bibfield
  {journal} {\bibinfo  {journal} {Phys. Rev. Lett.}\ }\textbf {\bibinfo
  {volume} {114}},\ \bibinfo {pages} {251602} (\bibinfo {year}
  {2015})}\BibitemShut {NoStop}%
\bibitem [{\citenamefont {Baggioli}\ \emph {et~al.}(2021)\citenamefont
  {Baggioli}, \citenamefont {Kim}, \citenamefont {Li},\ and\ \citenamefont
  {Li}}]{Baggioli:2021xuv}%
  \BibitemOpen
  \bibfield  {author} {\bibinfo {author} {\bibfnamefont {M.}~\bibnamefont
  {Baggioli}}, \bibinfo {author} {\bibfnamefont {K.-Y.}\ \bibnamefont {Kim}},
  \bibinfo {author} {\bibfnamefont {L.}~\bibnamefont {Li}}, \ and\ \bibinfo
  {author} {\bibfnamefont {W.-J.}\ \bibnamefont {Li}},\ }\href {\doibase
  10.1007/s11433-021-1681-8} {\bibfield  {journal} {\bibinfo  {journal} {Sci.
  China Phys. Mech. Astron.}\ }\textbf {\bibinfo {volume} {64}},\ \bibinfo
  {pages} {270001} (\bibinfo {year} {2021})},\ \Eprint
  {http://arxiv.org/abs/2101.01892} {arXiv:2101.01892 [hep-th]} \BibitemShut
  {NoStop}%
\bibitem [{\citenamefont {Amoretti}\ and\ \citenamefont
  {Musso}(2015)}]{Amoretti:2015gna}%
  \BibitemOpen
  \bibfield  {author} {\bibinfo {author} {\bibfnamefont {A.}~\bibnamefont
  {Amoretti}}\ and\ \bibinfo {author} {\bibfnamefont {D.}~\bibnamefont
  {Musso}},\ }\href {\doibase 10.1007/JHEP09(2015)094} {\bibfield  {journal}
  {\bibinfo  {journal} {JHEP}\ }\textbf {\bibinfo {volume} {09}},\ \bibinfo
  {pages} {094} (\bibinfo {year} {2015})},\ \Eprint
  {http://arxiv.org/abs/1502.02631} {arXiv:1502.02631 [hep-th]} \BibitemShut
  {NoStop}%
\bibitem [{\citenamefont {Blake}\ \emph {et~al.}(2015)\citenamefont {Blake},
  \citenamefont {Donos},\ and\ \citenamefont {Lohitsiri}}]{Blake:2015ina}%
  \BibitemOpen
  \bibfield  {author} {\bibinfo {author} {\bibfnamefont {M.}~\bibnamefont
  {Blake}}, \bibinfo {author} {\bibfnamefont {A.}~\bibnamefont {Donos}}, \ and\
  \bibinfo {author} {\bibfnamefont {N.}~\bibnamefont {Lohitsiri}},\ }\href
  {\doibase 10.1007/JHEP08(2015)124} {\bibfield  {journal} {\bibinfo  {journal}
  {JHEP}\ }\textbf {\bibinfo {volume} {08}},\ \bibinfo {pages} {124} (\bibinfo
  {year} {2015})},\ \Eprint {http://arxiv.org/abs/1502.03789} {arXiv:1502.03789
  [hep-th]} \BibitemShut {NoStop}%
\bibitem [{\citenamefont {Kittel}(2021)}]{kittel2021introduction}%
  \BibitemOpen
  \bibfield  {author} {\bibinfo {author} {\bibfnamefont {C.}~\bibnamefont
  {Kittel}},\ }\href@noop {} {\emph {\bibinfo {title} {Introduction to solid
  state physics Eighth edition}}}\ (\bibinfo {year} {2021})\BibitemShut
  {NoStop}%
\bibitem [{\citenamefont {Chagnet}\ and\ \citenamefont
  {Schalm}(2023)}]{Chagnet:2023xsl}%
  \BibitemOpen
  \bibfield  {author} {\bibinfo {author} {\bibfnamefont {N.}~\bibnamefont
  {Chagnet}}\ and\ \bibinfo {author} {\bibfnamefont {K.}~\bibnamefont
  {Schalm}},\ }\href@noop {} {\  (\bibinfo {year} {2023})},\ \Eprint
  {http://arxiv.org/abs/2303.17685} {arXiv:2303.17685 [cond-mat.str-el]}
  \BibitemShut {NoStop}%
\bibitem [{\citenamefont {Hayes}\ \emph {et~al.}(2016)\citenamefont {Hayes},
  \citenamefont {McDonald}, \citenamefont {Breznay}, \citenamefont {Helm},
  \citenamefont {Moll}, \citenamefont {Wartenbe}, \citenamefont {Shekhter},\
  and\ \citenamefont {Analytis}}]{Hayes2016}%
  \BibitemOpen
  \bibfield  {author} {\bibinfo {author} {\bibfnamefont {I.~M.}\ \bibnamefont
  {Hayes}}, \bibinfo {author} {\bibfnamefont {R.~D.}\ \bibnamefont {McDonald}},
  \bibinfo {author} {\bibfnamefont {N.~P.}\ \bibnamefont {Breznay}}, \bibinfo
  {author} {\bibfnamefont {T.}~\bibnamefont {Helm}}, \bibinfo {author}
  {\bibfnamefont {P.~J.~W.}\ \bibnamefont {Moll}}, \bibinfo {author}
  {\bibfnamefont {M.}~\bibnamefont {Wartenbe}}, \bibinfo {author}
  {\bibfnamefont {A.}~\bibnamefont {Shekhter}}, \ and\ \bibinfo {author}
  {\bibfnamefont {J.~G.}\ \bibnamefont {Analytis}},\ }\href {\doibase
  10.1038/nphys3773} {\bibfield  {journal} {\bibinfo  {journal} {Nature
  Physics}\ }\textbf {\bibinfo {volume} {12}},\ \bibinfo {pages} {916}
  (\bibinfo {year} {2016})}\BibitemShut {NoStop}%
\bibitem [{\citenamefont {Baggioli}\ and\ \citenamefont
  {Pujolas}(2017)}]{Baggioli:2016oqk}%
  \BibitemOpen
  \bibfield  {author} {\bibinfo {author} {\bibfnamefont {M.}~\bibnamefont
  {Baggioli}}\ and\ \bibinfo {author} {\bibfnamefont {O.}~\bibnamefont
  {Pujolas}},\ }\href {\doibase 10.1007/JHEP01(2017)040} {\bibfield  {journal}
  {\bibinfo  {journal} {JHEP}\ }\textbf {\bibinfo {volume} {01}},\ \bibinfo
  {pages} {040} (\bibinfo {year} {2017})},\ \Eprint
  {http://arxiv.org/abs/1601.07897} {arXiv:1601.07897 [hep-th]} \BibitemShut
  {NoStop}%
\bibitem [{\citenamefont {Gout\'eraux}\ \emph {et~al.}(2016)\citenamefont
  {Gout\'eraux}, \citenamefont {Kiritsis},\ and\ \citenamefont
  {Li}}]{Gouteraux:2016wxj}%
  \BibitemOpen
  \bibfield  {author} {\bibinfo {author} {\bibfnamefont {B.}~\bibnamefont
  {Gout\'eraux}}, \bibinfo {author} {\bibfnamefont {E.}~\bibnamefont
  {Kiritsis}}, \ and\ \bibinfo {author} {\bibfnamefont {W.-J.}\ \bibnamefont
  {Li}},\ }\href {\doibase 10.1007/JHEP04(2016)122} {\bibfield  {journal}
  {\bibinfo  {journal} {JHEP}\ }\textbf {\bibinfo {volume} {04}},\ \bibinfo
  {pages} {122} (\bibinfo {year} {2016})},\ \Eprint
  {http://arxiv.org/abs/1602.01067} {arXiv:1602.01067 [hep-th]} \BibitemShut
  {NoStop}%
\bibitem [{\citenamefont {Taylor}\ and\ \citenamefont
  {Woodhead}(2014)}]{Taylor:2014tka}%
  \BibitemOpen
  \bibfield  {author} {\bibinfo {author} {\bibfnamefont {M.}~\bibnamefont
  {Taylor}}\ and\ \bibinfo {author} {\bibfnamefont {W.}~\bibnamefont
  {Woodhead}},\ }\href {\doibase 10.1140/epjc/s10052-014-3176-9} {\bibfield
  {journal} {\bibinfo  {journal} {Eur. Phys. J. C}\ }\textbf {\bibinfo {volume}
  {74}},\ \bibinfo {pages} {3176} (\bibinfo {year} {2014})},\ \Eprint
  {http://arxiv.org/abs/1406.4870} {arXiv:1406.4870 [hep-th]} \BibitemShut
  {NoStop}%
\bibitem [{\citenamefont {Alberte}\ \emph {et~al.}(2016)\citenamefont
  {Alberte}, \citenamefont {Baggioli}, \citenamefont {Khmelnitsky},\ and\
  \citenamefont {Pujolas}}]{Alberte:2015isw}%
  \BibitemOpen
  \bibfield  {author} {\bibinfo {author} {\bibfnamefont {L.}~\bibnamefont
  {Alberte}}, \bibinfo {author} {\bibfnamefont {M.}~\bibnamefont {Baggioli}},
  \bibinfo {author} {\bibfnamefont {A.}~\bibnamefont {Khmelnitsky}}, \ and\
  \bibinfo {author} {\bibfnamefont {O.}~\bibnamefont {Pujolas}},\ }\href
  {\doibase 10.1007/JHEP02(2016)114} {\bibfield  {journal} {\bibinfo  {journal}
  {JHEP}\ }\textbf {\bibinfo {volume} {02}},\ \bibinfo {pages} {114} (\bibinfo
  {year} {2016})},\ \Eprint {http://arxiv.org/abs/1510.09089} {arXiv:1510.09089
  [hep-th]} \BibitemShut {NoStop}%
\bibitem [{\citenamefont {Davison}(2013)}]{Davison:2013jba}%
  \BibitemOpen
  \bibfield  {author} {\bibinfo {author} {\bibfnamefont {R.~A.}\ \bibnamefont
  {Davison}},\ }\href {\doibase 10.1103/PhysRevD.88.086003} {\bibfield
  {journal} {\bibinfo  {journal} {Phys. Rev. D}\ }\textbf {\bibinfo {volume}
  {88}},\ \bibinfo {pages} {086003} (\bibinfo {year} {2013})},\ \Eprint
  {http://arxiv.org/abs/1306.5792} {arXiv:1306.5792 [hep-th]} \BibitemShut
  {NoStop}%
\bibitem [{\citenamefont {Huang}\ \emph {et~al.}(2023)\citenamefont {Huang},
  \citenamefont {Sachdev},\ and\ \citenamefont {Lucas}}]{Huang:2023ihu}%
  \BibitemOpen
  \bibfield  {author} {\bibinfo {author} {\bibfnamefont {X.}~\bibnamefont
  {Huang}}, \bibinfo {author} {\bibfnamefont {S.}~\bibnamefont {Sachdev}}, \
  and\ \bibinfo {author} {\bibfnamefont {A.}~\bibnamefont {Lucas}},\
  }\href@noop {} {\  (\bibinfo {year} {2023})},\ \Eprint
  {http://arxiv.org/abs/2306.03130} {arXiv:2306.03130 [cond-mat.str-el]}
  \BibitemShut {NoStop}%
\bibitem [{\citenamefont {Amoretti}\ \emph
  {et~al.}(2020{\natexlab{b}})\citenamefont {Amoretti}, \citenamefont
  {Brattan}, \citenamefont {Magnoli},\ and\ \citenamefont
  {Scanavino}}]{Amoretti:2020mkp}%
  \BibitemOpen
  \bibfield  {author} {\bibinfo {author} {\bibfnamefont {A.}~\bibnamefont
  {Amoretti}}, \bibinfo {author} {\bibfnamefont {D.~K.}\ \bibnamefont
  {Brattan}}, \bibinfo {author} {\bibfnamefont {N.}~\bibnamefont {Magnoli}}, \
  and\ \bibinfo {author} {\bibfnamefont {M.}~\bibnamefont {Scanavino}},\ }\href
  {\doibase 10.1007/JHEP08(2020)097} {\bibfield  {journal} {\bibinfo  {journal}
  {JHEP}\ }\textbf {\bibinfo {volume} {08}},\ \bibinfo {pages} {097} (\bibinfo
  {year} {2020}{\natexlab{b}})},\ \Eprint {http://arxiv.org/abs/2005.09662}
  {arXiv:2005.09662 [hep-th]} \BibitemShut {NoStop}%
\bibitem [{\citenamefont {Else}(2023)}]{else2023holographic}%
  \BibitemOpen
  \bibfield  {author} {\bibinfo {author} {\bibfnamefont {D.~V.}\ \bibnamefont
  {Else}},\ }\href@noop {} {\enquote {\bibinfo {title} {Holographic models of
  non-fermi liquid metals revisited: an effective field theory approach},}\ }
  (\bibinfo {year} {2023}),\ \Eprint {http://arxiv.org/abs/2307.02526}
  {arXiv:2307.02526 [cond-mat.str-el]} \BibitemShut {NoStop}%
\end{thebibliography}%

\end{document}